\documentclass[12pt]{amsart}                                             
\usepackage{amssymb}
\usepackage{float}
\usepackage{verbatim}
\usepackage[usenames]{color}
\usepackage[all,cmtip]{xy}
\usepackage{tikz-cd}
\usepackage{hyperref}
\usepackage{url}
\usepackage[usenames]{color}
\usepackage{verbatim}
\usepackage[all]{xy}

\newtheorem{thm}{Theorem}
\newtheorem{prop}[thm]{Proposition} 
\newtheorem{lemma}[thm]{Lemma}

\theoremstyle{definition}
\newtheorem{df}[thm]{Definition}
\newtheorem{conv}[thm]{Convention}
\newtheorem{notat}[thm]{Notation}

\theoremstyle{remark}

\newtheorem{ex}[thm]{Example} 
\newtheorem{rmk}[thm]{Remark}
\newtheorem{ctr}[thm]{Category-Theoretic Remark}

\newenvironment{eqnum}{\begin{equation}}{\end{equation}}
\newenvironment{ls}{\begin{itemize}}{\end{itemize}}
\newenvironment{lsnum}{\begin{enumerate}}{\end{enumerate}}
\newenvironment{pf}{\begin{proof}}{\end{proof}}

\newcommand{\bld}[1]{\ensuremath{\mathbf {#1}}}
\newcommand{\bbb}[1]{\ensuremath{\mathbb {#1}}}
\newcommand{\h}{\bld}

\renewcommand{\phi}{\varphi}
\newcommand{\eps}{\varepsilon}
\newcommand{\sq}[1]{\ensuremath{\langle#1\rangle}}

\newcommand{\notarrow}{\kern .42em\not\kern -.42em\longrightarrow}
\renewcommand{\th}{\ensuremath{{}^{\text{th}}}}
\newcommand{\ket}[1]{\ensuremath{|#1\rangle}}

\newcommand{\w}[3]{\ensuremath{#1\vdash#2=#3}}
\newcommand{\x}{\ensuremath{\times}}

\newcommand{\one}[2]{(#1\underset{1}{\cdot}#2)}
\newcommand{\two}[2]{(#1\underset{\tau}{\cdot}#2)}

\newcommand{\ax}{\ensuremath{\alpha^\times}}
\newcommand{\ap}{\ensuremath{\alpha^{+}}}

\newcommand{\gp}{\ensuremath{\gamma^{+}}}
\newcommand{\gx}{\ensuremath{\gamma^\times}}

\renewcommand{\1}{^{-1}}

\newcommand\xqed[1]{%
  \leavevmode\unskip\penalty9999 \hbox{}\nobreak\hfill
  \quad\hbox{#1}}
\newcommand\textqed{\xqed{$\triangleleft$}}

\title{Witness Algebra and Anyon Braiding}
\author{Andreas Blass}
\address{Mathematics Department\\
University of Michigan\\
Ann Arbor, MI 48109--1043, U.S.A.}
\email{ablass@umich.edu}
\author{Yuri Gurevich}
\address{Computer Science and Engineering\\
University of Michigan\\
Ann Arbor, MI  48109-2121, U.S.A}
\email{gurevich@umich.edu}

\begin{document}

\begin{abstract}
Topological quantum computation employs two-dimensional quasiparticles called
anyons. The generally accepted mathematical basis for the theory of anyons is
the framework of modular tensor categories. That framework involves a
substantial amount of category theory and is, as a result, considered rather
difficult to understand. Is the complexity of the present framework necessary?
The computations of associativity and braiding matrices can be based on a much
simpler framework, which looks less like category theory and more like familiar
algebra. We introduce that framework here.
 \end{abstract}

\maketitle

\section{Introduction}          \label{sec:intro}

Topological quantum computation employs two-dimensional 
quasiparticles
called anyons \cite{Kitaev,FKLW}. The generally accepted mathematical
basis for the theory of anyons is the framework of modular
tensor categories. That framework, as presented in \cite{Wang} or
\cite{PP} or \cite{G225} involves a substantial amount of category
theory and is, as a result, considered rather difficult to
understand. For example, Trebst et al.\ \cite[page 385]{TTWL} write
``In general terms, we can describe anyons by a mathematical framework
called tensor category theory. $\dots$ Here we will not delve into
this difficult mathematical subject $\dots$ .'' Similarly, Bonderson
\cite[page~13]{Bonderson} writes ``In mathematical terminology, anyon
models are known as unitary braided tensor categories, but we will
avoid descending too far into the abstract depths of category theory
$\dots$.''

Is the complexity of the current framework necessary? We do not think
so. Our opinion is based on the following.  
\begin{enumerate}
\item Our own experience, admittedly modest. In \cite{G225}, after
  describing modular tensor categories, we presented a simplification
  based on Yoneda's Lemma. Then we exhibited some computations for the 
  particular case of Fibonacci anyons, including the computation of
  the braiding operators that are central to proposed uses of anyons
  for quantum computation.  It turned out that a good part of the
  axiomatics of tensor categories was not needed for those
  computations.
\item Only isomorphisms, rather than category theory's more general
  morphisms, are used in the anyon theory.
\item Physicists tend to avoid category theory.
\end{enumerate}
The computations in \cite{G225} can be based on a much simpler
framework, which looks less like category theory and more like
familiar algebra.   We introduce that framework here. 

The main idea of the proposed framework is to work with ordinary
algebraic structures, like rings, amplified with a notion of
\emph{witnesses} for equations.  The rules of equational logic are
then accompanied by constructions of new witnesses from old.  For
example, where equational logic says that, from an equation $a=b$, one
can infer $b=a$, our framework will say that, from a witness for
$a=b$, one can construct a witness for $b=a$.  The other rules of
equational logic are treated similarly; see Section~\ref{sec:logic}
for details.  We call this new framework \emph{witness algebra}.

Traditional equational logic can be viewed as the special case of
witness algebra where all witnesses are trivial and therefore do not
need to be mentioned.  In our work with anyons, the witnesses for the
associativity and commutativity of multiplication will be highly
nontrivial.  Other witnesses, for example those for the associativity
and commutativity of addition, will not be entirely trivial but will
amount to minor bookkeeping information.  See Section~\ref{sec:fusion}
for details.

\begin{ctr}
Readers who care only about the witness algebra framework and not the
traditional category-theoretic one can skip this and subsequent
category-theoretic remarks. These will serve to connect our framework
to the traditional one but will not be essential for the development
of witness algebra itself.  

Our work, having emerged from category theory, can be translated back
into category-theoretic terminology.  The result of that translation
would be a theory of groupoids (categories in which all morphisms are
isomorphisms) enhanced with additional structure.  For example, in our
braid semirings (defined in Section~\ref{sec:braid}), the enhancement
would be two monoidal structures, one of which (corresponding to
addition) is symmetric while the other (corresponding to
multiplication) is braided, plus a distributive law connecting the
two, plus suitable coherence axioms. \textqed
\end{ctr}

Somewhat ironically, easing the category-theoretical complexity of the
anyon theory required some category-theoretical investigation. The
associativity, commutativity and distributivity isomorphisms should
satisfy appropriate coherence conditions. What are these conditions?
Much of the work has been done in the literature but there was a gap
which we filled in the companion paper \cite{cohere}. The situation is
explained in detail in Section~\ref{sec:braid}. 

\subsection*{Related Work} G\"{o}del suggested in unpublished work
\cite[Section~V]{Goedel} that modal logic could profitably be
amplified by introducing justifications. This
idea was independently rediscovered and extensively studied by
Artemov; see for example \cite{Artemov}.  The role of justifications
in modal logic is similar to the role of witnesses in our witness
algebra.  But, instead of modal logic, we work with elementary
algebra. Another difference is that Artemov's justifications can be
nested; that is, ``$\xi$ is a justification for $\phi$'' is itself a
formula that can admit justifications.  We use witnesses only for
equations, and ``$\xi$ is a witness for $a=b$'' is not itself an
equation, so it cannot be witnessed.

\section{Witnessed Equational Logic}    \label{sec:logic}

\subsection{Preliminaries}
To establish terminology, we recall some definitions from universal
algebra.  A \emph{signature} is a collection of operation symbols,
each having a specified number of argument places (arity).  An
\emph{algebra} $A$ of signature $\Sigma$ consists of a set $|A|$, the
\emph{universe} of $A$, together with interpretations of the symbols
in $\Sigma$ on $|A|$; an $r$-ary symbol is interpreted as an $r$-ary
operation.  The interpretations of nullary operation symbols are
viewed as elements of $|A|$; that is, nullary operation symbols are
individual constants.

It is often convenient to use the same symbol for an algebra and its
universe; in such cases it should be clear from the context whether
the symbol denotes the algebra or its universe. 

\subsection{Witness Frames}
We begin the discussion of witness algebras with the special case
where the signature $\Sigma$ is empty. In this case, algebras of
signature $\Sigma$ are merely sets with no additional
structure. Nevertheless, the key concept of witnessing appears even in
this context and is easier to explain without the additional baggage
of a nontrivial signature. So we begin with this special case; we use
the name ``witness frame'' for witness algebras for the empty
signature.\footnote{Since algebras for the empty signature are sets,
  it would seem reasonable to call witness algebras for the empty
  signature ``witness sets''.  Unfortunately, that sounds too much
  like a set of witnesses. We have chosen the terminology ``witness
  frame'' to suggest that these are basic systems to which additional
  algebraic structure (nonempty signatures) could be attached.}

The idea here is that we have a set $S$, each of whose elements $s$
might be viewed in a variety of ways.  Witnesses will indicate whether
and how two views represent the same element $s$.  As a fairly typical
example (similar to what we shall later use for anyon computations),
the elements of $S$ might be some vector spaces, and a view of a
vector space might be that space equipped with a particular basis.  A
witness should then indicate how two vector spaces with specified
bases (i.e., two views) are actually the same vector space (though the
bases might be quite different). Such a witness could be an invertible
matrix transforming the one basis into the other, i.e., expressing
each vector of the latter basis as a linear combination of the former
basis.

Thus, a witness frame really involves two sets: the set $S$ and the set
of views of elements of $S$. To formalize these notions, it is
convenient to take the views as the basic entities; elements of $S$
can then be identified with equivalence classes under the equivalence
relation of ``some witness shows that the two views represent the same
element of $S$.'' (Of course, the formal definition will need to
ensure, among other things, that this is an equivalence relation.)

This discussion leads to the somewhat unusual situation that the
entities of primary interest and importance, the elements of $S$, are
not the ones taken to be basic in the formalization.  We shall
emphasize the importance of the equivalence classes, the elements of
$S$, by the following terminology and notation.

The views, which are formally the basic entities, will be called the
raw elements of the witness frame, and the symbol $\equiv$ will be used
to mean that two of them are the same
view, not merely representatives of the
same element of $S$. The equivalence relation of ``some witness says
they're the same'' will be denoted by $=$, and the equivalence
classes, the elements of $S$, will be called the true elements of the
witness frame.  In the vector space example above, $v\equiv w$ would
mean that the views $v$ and $w$ are the same vector space with the
same basis, whereas $v=w$ would mean that they are the same vector
space with possibly different bases.

\begin{rmk}
Our use of $=$ to denote an equivalence relation coarser than what one
might consider complete equality, $\equiv$, is unusual but not
unprecedented. For example, in combinatorial group theory, when
discussing groups and their presentations, one sometimes writes $v=w$ to
mean that the words $v$ and $w$ represent the same group element, while
$v\equiv w$ means that they are identical as words.\textqed
\end{rmk}

We emphasize that our unusual use of the equality symbol applies only
to raw elements. In other contexts, the equality symbol retains its
customary meaning. In particular, if $\xi$ and $\eta$ are witnesses,
then $\xi=\eta$ means that they are the same witness.

After these explanations of the underlying intention, we are ready for
the definition of witness frames.

\begin{df}              \label{frame}
  A \emph{witness frame}
  consists of a set $A$ of \emph{raw elements} and, for all elements
  $a,b\in A$, pairwise disjoint (possibly empty) sets $W(a,b)$ of
  \emph{witnesses} for equality between $a$ and $b$.  If
  $\xi\in W(a,b)$, we write \w\xi ab and we say that $\xi$
  \emph{witnesses} that $a=b$.  We call $\xi$ a \emph{witness} if it
  is in one of the sets $W(a,b)$. The system of raw elements and
  witnesses is required to have the following structure.
  \begin{lsnum}
    \item[(W1)] For each $a\in A$, there is a specified witness \w{1_a}aa.
    \item[(W2)] For each witness \w\xi ab, there is a specified
      \w{\xi^{-1}}ba.
\item[(W3)] For each pair of witnesses \w\xi ab and \w\eta bc, there is a
  specified witness \w{\xi*\eta}ac.
  \end{lsnum}
  These specifications are subject to the following axioms.
  \begin{lsnum}
    \item[(W4)] If \w\xi ab then $1_a*\xi=\xi*1_b=\xi$.
\item[(W5)] If \w\xi ab then $\xi*\xi^{-1}=1_a$ and
  $\xi^{-1}*\xi=1_b$. 
\item[(W6)] If \w\xi ab, \w\eta bc, and \w\zeta cd,\\ then
  $(\xi*\eta)*\zeta =\xi*(\eta*\zeta)$. \textqed
  \end{lsnum}
\end{df}

If we ignore the witnesses and pay attention only to the equations,
then requirements (W1), (W2), and (W3) correspond to the usual axioms
and rules of equational logic (in the case of the empty signature),
saying that equality is reflexive, symmetric, and transitive. Thus, in
witness frames these three requirements ensure that the relation $=$
introduced in the following definition is an equivalence relation.

\begin{df}
  In any witness frame, with the notation of the preceding definition,
  we define, for $a,b\in A$,
\[
a=b\iff(\exists\xi)\,\w\xi ab.
\]
The elements of $A$ are called \emph{raw elements} of the witness frame,
and equality between them is symbolized by $\equiv$. The equivalence
classes with respect to the relation $=$ just defined are called
\emph{true elements} of the witness frame. \textqed
\end{df}

We have built into the definition of witness frames that the sets
$W(a,b)$ for the various pairs $a,b$ are disjoint. In other words, a
witness $\xi$ witnesses only a single equation.  This convention is
very convenient for theoretical purposes.  It implies in particular
that the inverse $\xi^{-1}$ of any witness $\xi$ is uniquely defined and that the composition $\xi*\eta$ of any two witnesses $\xi, \eta$ can be defined in at most one way. In some concrete
situations, on the other hand, it is tempting to re-use the same
witness for several equations. Indeed, this temptation arose in our
vector space example above; one and the same matrix can serve as the
transformation matrix between many pairs of bases. Fortunately, there
is an easy solution for this problem, namely to ``mark'' each witness
with the equation that we want it to witness. That is, if the same
$\xi$ is in both $W(a,b)$ and $W(c,d)$, we replace it by \sq{a,\xi,b}
as an element of $W(a,b)$, and we replace it by \sq{c,\xi,d} as an
element of $W(c,d)$.  More generally, we adopt the following
convention, intended to give us the best of both worlds ---
authorization to re-use witnesses when describing concrete examples
while maintaining disjointness of the sets of witnesses for official
purposes.

\begin{conv}            \label{disjoint}
  If a witness frame is described in a way that allows the sets $W(a,b)$
  to overlap, then it is to be understood that the actual, official witnesses
  for an equation $a=b$ are not the described $\xi$'s but rather the
  marked versions \sq{a,\xi,b}. \textqed
\end{conv}

Our definition of witness frames includes requirements (W4), (W5), and
(W6), which say that certain combinations of witnesses are equal.  In
each case, the required equality makes sense because the two sides are
witnesses of the same equation. The intention behind these
requirements is that the specifications in (W1), (W2), and (W3) should
not be made randomly or arbitrarily but in some coherent way. For
example, the witness $\xi*\eta$ in (W3) should not be just any witness
for $a=c$ but rather one that combines, in a sensible way, the
information in the witnesses $\xi$ and $\eta$ (and the transitive law
of equality).

\begin{ctr}
  Category theorists will recognize witness frames as just a notational
  variant of the familiar notion of groupoid, a category in which all
  morphisms are isomorphisms. Our raw elements are the objects of the
  groupoid, our witnesses are the morphisms, and \w\xi ab amounts to
  $\xi:a\to b$. Our (W1), (W3), (W4), and (W6) amount to the
  definition of a category, while (W2) and (W5) provide the inverses
  that make all the morphisms isomorphisms. The true elements are the
  connected components of the groupoid. \textqed

Notice that we compose witnesses in what is often called diagrammatic
order. That is, the composition of \w\xi ab with \w\eta bc is written
as $\xi*\eta$, not as $\eta*\xi$ or as $\eta\circ\xi$.

We emphasize that these groupoids are quite different from the
abelian categories used in \cite{PP} and \cite{G225}. It is easy to
check that the only way an abelian category can be a groupoid is to be
trivial, i.e., to be equivalent to the category consisting of just a
single object and its identity morphism. \textqed
\end{ctr}

The following lemma records some basic properties of the operations on
witnesses that are involved in witness frames.

\begin{lemma}   \label{inv-cancel}
\mbox{}
  \begin{lsnum}
    \item The operation $*$ on witnesses admits cancellation. That is,
      if $\xi*\eta$ and $\xi*\zeta$ are defined and equal, then
      $\eta=\zeta$. Similarly if $\eta*\xi=\zeta*\xi$.
    \item Inversion is involutive. That is, all witnesses $\xi$
      satisfy $(\xi^{-1})^{-1}=\xi$.
  \end{lsnum}
\end{lemma}

\begin{pf}
For part~(1), assume $\xi*\eta=\xi*\zeta$, where $\w\xi ab$ and
$\w{\eta,\zeta}bc$ so that the $*$ operation is defined for these
witnesses. Then compute 
\begin{multline*}
\eta=1_b*\eta=(\xi^{-1}*\xi)*\eta=\xi^{-1}*(\xi*\eta)  =\\
\xi^{-1}*(\xi*\zeta)=(\xi^{-1}*\xi)*\zeta=1_b*\zeta=\zeta.
\end{multline*}

The proof under the hypothesis $\eta*\xi=\zeta*\xi$ is symmetrical. 

For part~(2), notice that, if $\w\xi ab$ then
$(\xi^{-1})^{-1}*\xi^{-1}=1_a=\xi*\xi^{-1}$, and use part~(1) to
cancel $\xi^{-1}$.
\end{pf}

\begin{rmk}
  The definition of witness frames requires that the raw elements, and
  therefore also the true elements, constitute a set rather than a
  proper class. Everything we do, however, would work just as well if
  we used classes instead. So, for example,we could deal with a
  witness frame whose true elements are all of the vector spaces,
  not just some. Except for this remark, we shall ignore the set-class
  distinction in this paper. \textqed
\end{rmk}

\subsection{Witness Algebras}
In the preceding subsection, we dealt with the case of the empty
signature, where algebras are just sets.  In keeping with that special
case, although our witness frames had considerable structure, as
described in (W1)--(W6), the true elements formed just a set with
no additional structure.  In the present subsection, we deal with the
case of general signatures $\Sigma$.  A witness $\Sigma$-algebra will
have enough additional structure to make the true elements into a
$\Sigma$-algebra in the usual sense.

\begin{df}\label{wa}
  Let $\Sigma$ be a signature. A \emph{witness}
  $\Sigma$-\emph{algebra} is a witness frame $A$ (with notation as
  above) together with actions of the operation symbols from $\Sigma$
  on raw elements and on witnesses, as follows.  Let $f\in\Sigma$ be
  $r$-ary. Then:
\begin{lsnum}
  \item $f:A^r\to A$ is an $r$-ary operation on the raw elements.
\item If $\w{\xi_i}{a_i}{b_i}$ for $i=1,\dots,r$ then
  $f(\xi_1,\dots,\xi_r)$ is defined and 
\[
\w{f(\xi_1,\dots,\xi_r)}{f(a_1,\dots,a_r)}{f(b_1,\dots,b_r)}.
\]
\item If $\w{\xi_i}{a_i}{b_i}$ and $\w{\eta_i}{b_i}{c_i}$ for
  $i=1,\dots,r$ then
\[\phantom{mmmmn}
f(\xi_1,\dots,\xi_r)*f(\eta_1,\dots,\eta_r)=
f(\xi_1*\eta_1,\dots,\xi_r*\eta_r). 
\phantom{mmm} \textqed \]
\end{lsnum}
\end{df}

To avoid a proliferation of notation, we use $f$ for all three of a
symbol in $\Sigma$, its action as an operation on $A$, and its action
on witnesses (as long as the context prevents ambiguity).

When the arity $r$ is zero, a 0-ary operation on $A$ is, of course, a
function from the one-element set $A^0$ into $A$. It is convenient
(and customary in algebra and logic) to identify such a function with
its unique value. Thus, 0-ary operations amount to constants.

If we pay attention only to witnessed equality and not to the specific
witnesses, then clause~(2) in Definition~\ref{wa} says that the
algebraic laws of equality concerning function symbols are obeyed: we
can substitute equals for equals. This clause and the clauses in the
earlier definition of witness frames give all the usual laws of
equational logic.

In the definition of witness algebras, clause~(1) makes the raw
elements into a $\Sigma$-algebra, which we call the \emph{raw
  algebra}. Clause~(2) makes the witnessed equality relation $=$,
which we already know is an equivalence relation, into a congruence
relation. As a result, the quotient, the set of true elements, becomes
a $\Sigma$-algebra as well.  We call it the \emph{true algebra} (or
the \emph{true $\Sigma$-algebra) }of the
witness algebra.

The purpose of clause~(3) is, as with some of the clauses in the
definition of witness frames earlier, to require that the witnesses
$f(\xi_1,\dots,\xi_r)$ in clause (2) should not be just arbitrarily
chosen witnesses for the relevant equations
$f(a_1,\dots,a_r)=f(b_1,\dots,b_r)$ but should be chosen in some
reasonable way based on $\xi_1,\dots,\xi_r$.

\begin{ex}     \label{trivial}
  Any $\Sigma$-algebra $A$ can be converted trivially into a witness
  algebra whose raw and true algebras are both the given $A$. Just
  take all the sets $W(a,b)$ to be singletons.  (We could even choose
  to take just a single, trivial witness, say 0, to witness exactly
  those equalities $a=b$ where $a$ and $b$ are the same raw element.
  Convention~\ref{disjoint} would make this choice ``legal'' by
  replacing the single witness 0 with different witnesses for
  different equations.) In this way, we can regard ordinary
  algebras as a special case of witness algebras.\textqed
 \end{ex}

 \begin{ex}             \label{quot}
 More generally, consider a $\Sigma$-algebra $A$, a
  congruence relation $\sim$ on it, and the quotient algebra
  $A/\!\!\sim$.  Then, using only one witness $0$ (until
  Convention~\ref{disjoint} turns it into many witnesses), we can produce
  a witness algebra whose raw algebra is $A$ and whose true algebra
  is $A/\!\!\sim$.  It suffices to define $0\vdash a=b$ to hold if and
  only if $a\sim b$.\textqed
 \end{ex}

As already mentioned in connection with some of the clauses in our
definitions, our witnesses behave similarly to deductions in equational
logic. To emphasize the similarity, we can use a notation where a
witness is displayed above the equation that it witnesses, resembling a
deduction of that equation.
Then we have, for all $a\in A$,
\[
  \begin{matrix}
    1_a\\ a= a,
  \end{matrix}
\]
corresponding to the axiom $a=a$ of equational logic.  For any 
\[
  \begin{matrix}
    \xi \\ a= b,
  \end{matrix}
\]
we can depict $\xi\1$ as
\[
  \begin{matrix}
    \xi \\ a= b \\
\cline{1-1}
b= a,
  \end{matrix}
\]
so that it looks like $\xi$ followed by an application of the
inference rule ``from $a=b$ infer $b=a$.'' Similarly, for any
\[
  \begin{matrix}
    \xi \\ a= b
  \end{matrix}
\text{\qquad and \qquad}
\begin{matrix}
  \eta \\ b= c,
\end{matrix}
\]
we can depict $\xi\ast\eta$ as 
\[
  \begin{matrix}
\xi &&\eta\\
a= b && b= c\\
\cline{1-3}
&a= c,&
      \end{matrix}
\]
so that it looks like $\xi$ and $\eta$ followed by an application of
the inference rule ``from $a=b$ and $b=c$ infer $a=c$.''  Finally, for
an $n$-ary function symbol $f$, we can depict $f(\xi_1,\dots,\xi_r)$
as
\begin{center}
$\displaystyle
  \begin{matrix}
    \xi_1 &\dots&\xi_r\\
a_1= b_1&\dots& a_r= b_r\\
\cline{1-3}  
&f(a_1,\dots,a_r)= f(b_1,\dots,b_r)   
  \end{matrix}
$\end{center} \vspace{-4ex}\textqed

\vspace{4ex}
The similarity between witnesses and deductions suggests the following
variation on the theme of Example~\ref{quot}.

\begin{ex}              \label{quot2}
Suppose $A$ is the $\Sigma$-algebra presented
  by some generators and relations, and suppose $B$ is the quotient
  obtained by imposing some additional relations.  Then we can almost
  obtain a witness algebra by taking $A$ as the raw algebra, taking
  the witnesses for any equation $a= b$ to be the formal deductions of
  this equation from the relations defining $B$, and taking witnesses
  $1_a,\xi\1,\xi\ast\eta$, and $f(\bar\xi)$ to be compositions of
  deductions as depicted above.  The reason for ``almost'' in the
  preceding sentence is that the axioms for witness algebras require
  certain identifications between deductions.  For example, two
  consecutive uses of symmetry as in
\[
  \begin{matrix}
    \xi\\a= b\\ \cline{1-1}b= a\\ \cline{1-1}a= b
  \end{matrix}
\]
should have no effect (see part~(2) of Lemma~\ref{inv-cancel});
this deduction should be identified with  
\[
  \begin{matrix}
\xi\\a= b.    
  \end{matrix}
\]
A special case of this example occurs when $A$ is the algebra in a
certain variety presented by certain generators and relations, and
$B$ is presented by the same generators and relations in a smaller
variety, given by more identities.  For example, $A$ might be the
group with a certain presentation while $B$ is the abelian group
with the same presentation (the abelianization of $A$).  The
witnesses would then be deductions (modulo identifications required by
the axioms) from the commutative law.\textqed
\end{ex}

\begin{ctr}
Since witness algebras amount to groupoids, our definition of witness
$\Sigma$-algebras makes the operations $f\in\Sigma$ act on both the
objects (raw elements) and morphisms (witnesses). These operations 
constitute functors from powers of  the groupoid to itself. In
general, the definition of ``functor'' also requires preservation of
identity morphisms, so it would seem that we need to require
\[
f(1_{a_1},\dots,1_{a_r})= 1_{f(a_1,\dots,a_r)}.
\]
In the case of groupoids, though, this preservation of identity
elements follows from compatibility with composition. This last
obsrevation will be useful even apart from the category-theoretic
point of view, so we formulate it as the following
proposition. 
\end{ctr}

\begin{prop}    \label{elem}\mbox{}
If $f\in\Sigma$ is $r$-ary then 
$f(1_{a_1},\dots,1_{a_r}) = 1_{f(a_1,\dots,a_r)}$.
In particular, if $f$ is a nullary operation symbol, then the corresponding witness is $1_f$.
\end{prop}

\begin{pf}
To simplify the notation, we assume for now that $f$ is binary, and we
write $a$ and $b$ instead of $a_1$ and $a_2$. Note that both
$f(1_a,1_b)$ and $1_{f(a,b)}$ witness the equation
$f(a,b)=f(a,b)$. Furthermore, according to requirement~(3) in the
definition of witness algebras, we have 
\[
 f(1_a,1_b) \ast f(1_a,1_b) = f(1_a\ast1_a,1_b\ast1_b) = f(1_a,1_b) =
 f(1_a,1_b)\ast 1_{f(a,b)}.
\]
The desired conclusion now follows by cancellation (part~(1) of
Lemma~\ref{inv-cancel}).   

The argument for $r=2$ generalizes easily to larger arities $r$ and to
$r=1$. For $r=0$, the argument still works and in fact becomes simpler,
but it requires some notational caution, as follows. If $f$ is 0-ary
then, remembering that 0-ary operations on a set amount to elements, we
have a raw element $f$, and we also have a witness $\tilde f$ (also
called $f$ if no confusion results, but here confusion would result) for
the equation $f=f$. Then we have, from the definition of witness
algebra,
\[
\tilde f*\tilde f=\tilde f=\tilde f*1_f,
\]
and cancellation (part~(1) of Lemma~\ref{inv-cancel}) gives us $\tilde
f=1_f$.
\end{pf}

\section{Braid Semirings}                    \label{sec:braid}

In this section, we introduce the particular algebraic theory
underlying the application of witness algebra to anyons.  The basic
idea is quite simple, but various complications arise in the
details.

We shall arrange our witness algebras so that the associated true
algebras are commutative semirings with unit.  Here ``semiring'' is
defined like ``ring'' except that additive inverses are not required
to exist.  Some authors use ``rig'' to mean ``semiring'' (the idea
being that removing the letter ``n'' from ``ring'' corresponds to
removing \textbf{n}egatives from the definition). Because
multiplication will be commutative in our semirings, we adopt the
following convention for brevity.

\begin{conv}            \label{rig}
  ``Rig'' means commutative semiring with unit.
\end{conv}

Thus, the signature for our algebras consists of two binary operations
$+$ and $\times$ and two constants $0$ and $1$.  A rig is an algebra
for this signature in which
\begin{ls}
  \item both $+$ and $\times$ are associative and commutative,
\item $0$ and $1$ are identity elements for $+$ and $\times$,
  respectively, and
\item $\times$ distributes over $+$.
\end{ls}
Distributivity here means not only that multiplication distributes
over sums of two elements\footnote{We begin here to use capital
  letters for elements of our raw algebras, for two reasons. First, in
the application to anyons, these raw elements will be tuples of vector
spaces, for which capital letters are more natural than lower-case
letters. Second, we shall need to import some diagrams from
\cite{cohere}. That paper was written from the point of view of
category theory, so raw elements were objects of categories, denoted,
as usual, by capital letters.}, $A\times(B+C)=(A\times B)+(A\times C)$
(which implies distribution over any larger number of summands) but
also distribution over zero summands, i.e., $A\times0=0$. (This
equation would be redundant in the presence of additive inverses, but
it needs to  be assumed in the axiomatization of rigs.)

The raw algebras admit an even simpler description: They are algebras
for the same signature $\{+,\times,0,1\}$ but subject to no equations.
This means that, for the true algebras to satisfy the equations that
define rigs, we need to introduce witnesses for all those equations.
That is, our witness algebras must have (at least) the following
witnesses specified as part of the structure, for all raw elements
$A,B,C$:
\begin{align*}
&\mathbf{associative}\ + 
&& \w{\alpha^{+}_{A,B,C}}{(A+ B)+ C}{A+(B+ C)}\\
&\mathbf{unit}\ + 
&&\w{\lambda^+_A}{0+ A}{A}\quad\text{ and}\quad \w{\rho^+_A}{A+0}A\\
&\mathbf{commutative}\ +
&&\w{\gamma^+_{A,B}}{A+ B}{B+ A}\\
&\mathbf{associative}\ \times 
&&\w{\alpha^{\times}_{A,B,C}}{(A\times B)\times C}{A\times(B\times C)}\\
&\mathbf{unit}\ \times
&&\w{\lambda^\times_A}{1\times A}A\quad\text{ and}\quad
\w{\rho^\times_A}{A\times1}A\\
&\mathbf{commutative}\ \times
&&\w{\gamma^\times_{A,B}}{A\times B}{B\times A}\\
&\mathbf{distributive}\ 2
&&\w{\delta_{A,B,C}}{A\times(B+ C)}{(A\times B)+(A\times C)} \\
&\mathbf{distributive}\ 0
&&\w{\eps_A}{A\times 0}0
\end{align*}
We shall refer to these witnesses as the \emph{rig witnesses}.

So far, our description of the desired witness algebras can be easily
summarized: Algebras for the signature $\{+,\times,0,1\}$ with enough
specified witnesses to make the true algebra a rig.  More is needed,
though, for this structure to make good sense and (more importantly)
to be useful for anyon theory. We need to specify, or at least
constrain, how the rig witnesses listed here interact with each other
and with other witnesses that might be present.  Let us begin with two
examples before presenting the general situation.

\begin{ex} \label{nat-gamma-plus} Suppose we have witnesses
  $\w\xi AA'$ and $\w\eta BB'$. Then the fact that $A$ and $B$ commute
  under addition has, in addition to the witness $\gamma^+_{A,B}$
  above, another witness that works via the commutativity of $A'$ and
  $B'$, namely $(\xi+\eta)*\gamma^+_{A',B'}*(\eta+\xi)^{-1}$.  If we
  think of $\xi$ and $\eta$ as giving us alternative ways to view $A$
  and $B$, then this second witness for $A+B=B+A$ is just an
  alternative way to view the original witness $\gamma^+_{A,B}$.  So
  it is reasonable to require that these two witnesses be
  equal\footnote{Recall that equality of witnesses means genuine
    identity, not existence of some meta-witness.}.  Equivalently,
\[
\gamma^+_{A,B}*(\eta+\xi)=(\xi+\eta)*\gamma^+_{A',B'}.
\]
This requirement can be summarized as ``The witnesses $\gamma^+$
respect witnessed equalities.''  In Section~\ref{nat}, we will impose
analogous requirements on all of the other rig witnesses. The
justifications for these requirements are the same as for $\gamma^+$
here. \textqed
\end{ex}

\begin{ctr}
In category theory, the requirement in Example~\ref{nat-gamma-plus} is
expressed by saying that $\gamma^+$ is a natural transformation.  The
analogous requirements for the other rig witnesses will say that all
of them are natural transformations. 
\end{ctr}

\begin{ex}      \label{coh-comm-zero}
We have two witnesses for $A+0=A$, namely $\rho^+_A$ and
$\gamma^+_{A,0}*\lambda^+_A$.  It seems reasonable to require that
they coincide.  \textqed
\end{ex}

The requirements in Example~\ref{nat-gamma-plus} and
Example~\ref{coh-comm-zero} are qualitatively different. In the
former, we were concerned with how a rig witness ($\gamma^+$)
interacts with arbitrary other witnesses ($\xi$ and $\eta$); in the
latter, we are concerned with how rig witnesses interact with each
other.  In the former, there is an obvious generalization from the
$\gamma^+$ considered there to all the other rig witnesses. In the
latter, there is no obvious generalization, and indeed there is
considerable freedom in choosing what requirements of this sort should
be imposed.

The rest of this section is devoted to presenting the requirements
that we impose on the witnessess in our braid rigs (also called braid
semirings).  We present these requirements in three parts,
subsections~\ref{nat}, \ref{coh-mon}, and \ref{coh-distrib}, with an
intermediate subsection~\ref{coh-criteria} explaining how some of the
requirements were chosen.  

The requirements in subsection~\ref{nat}
are analogous to what we saw in Example~\ref{nat-gamma-plus}. This
subsection is simply the evident generalization of the example from
$\gamma^+$ to all rig witnesses.  Subsections~\ref{coh-mon} and
\ref{coh-distrib} present requirements analogous to that in
Example~\ref{coh-comm-zero}.  As indicated above, we have some freedom
in choosing requirements of this sort. Subsection~\ref{coh-criteria}
discusses how we chose to exercise that freedom in making the decisions
in the next two subsections.

Much of the material in subsections~\ref{coh-criteria},
\ref{coh-mon}, and \ref{coh-distrib} has already appeared, up to
  notational and terminological differences, in our earlier paper
  \cite{cohere}.  For the reader's convenience, we import it here,
  with the necessary modifications, rather than contenting ourselves
  with a list of the changes.
  
Braid rigs will be defined as witness algebras for the rig signature
$\{+,\times,0,1\}$, equipped with all the rig witnesses listed above
and satisfying all the requirements imposed in the following
subsections. 

\subsection{Rig Witnesses Respect Witnessed
  Equalities}         \label{nat}   
Suppose we have witnesses 
\[
\w\xi A{A'},\quad\w\eta B{B'}, \quad\text{and}\quad\w\zeta C{C'}.
\]
Then we require that all rig witnesses involving any of $A,B,C$ and
the corresponding rig witnesses involving $A',B',C'$ match via the
given $\xi,\eta,\zeta$.  In detail, these requirements are as follows.
\begin{align*}
  \alpha^+_{A,B,C}*(\xi+(\eta+\zeta))
  &=((\xi+\eta)+\zeta)*\alpha^+_{A',B',C'}\\ 
\lambda^+_A*\xi&=(1_0+\xi)*\lambda^+_{A'}\\
\rho^+_A*\xi&=(\xi+1_0)*\rho^+_{A'}\\
\gamma^+_{A,B}*(\eta+\xi)&=(\xi+\eta)*\gamma^+_{A',B'}\\
\alpha^\times_{A,B,C}*(\xi\times(\eta\times\zeta))
&=((\xi\times\eta)\times\zeta)*\alpha^\times_{A',B',C'}\\
\lambda^\times_A*\xi&=(1_1\times\xi)*\lambda^\times_{A'}\\
\rho^\times_A*\xi&=(\xi\times1_1)\times\rho^\times_{A'}\\
\gamma^\times_{A,B}*(\eta\times\xi)&=
(\xi\times\eta)*\gamma^\times_{A',B'}\\
\delta_{A,B,C}*((\xi\times\eta)+(\xi\times\zeta))&=
(\xi\times(\eta+\zeta))*\delta_{A',B',C'}\\
\eps_A&=(\xi\times1_0)*\eps_{A'}
\end{align*}

Notice that the fourth of these equations is what we had in
Example~\ref{nat-gamma-plus}.  All ten of the equations have the same
general form: The first factor on the left and the second on the right
are from one of our ten types of rig witnesses, with unprimed
subscripts on the left and primed on the right. The other factors are
built from (some of) $\xi,\eta,\zeta$.  How they are built matches the
specifications of the rig witness.  (In the last of these equations,
in strict analogy to the previous ones, the left side would have been
$\eps_A*1_0$; we have performed the trivial simplification to
$\eps_A$.)

\begin{ctr}
  In the language of category theory, these ten equations merely say
  that the rig witnesses constitute ten natural
  transformations. \textqed 
\end{ctr}

\subsection{Coherence Conditions} \label{coh-criteria} 

We turn next to identities, known as \emph{coherence conditions},
between various compositions of rig witnesses. We have seen one
coherence condition, $\gamma^+_{A,0}*\lambda^+_A=\rho^+_A$, in
Example~\ref{coh-comm-zero}, but there is no evident way to determine
all of the coherence conditions that should be imposed on our rig
witnesses.  In fact, as we shall see, the choice of coherence
conditions is influenced by the intended application.
For example, since addition alone and multiplication alone are subject
to the same requirements in the definition of rig (associativity,
unit, and commutativity), it would seem natural to impose the same
coherence conditions on the purely additive and purely multiplicative
rig witnesses. That approach, however, turns out to be completely
inappropriate for the study of non-abelian anyons. (See the discussion
of symmetry versus braiding later in the present subsection.)
Accordingly, we devote this subsection to explaining how we
selected suitable coherence conditions to be satisfied by braid rigs;
the conditions themselves will be presented in the next two
subsections.

There are two mathematical constraints on our selection of coherence
conditions, plus a practical consideration that also influenced our
choices.  The first and most important mathematical constraint is that
our coherence conditions should be satisfied in the witness algebras
arising in anyon models, the witness algebras that we propose as a
simplification of modular tensor categories.  Axioms that fail in the
intended examples are useless.  So we must not make our coherence
conditions too strong.

The second mathematical constraint is that our coherence conditions
should not be too weak; they should entail all the information needed
in our computations of specific examples. For example, our coherence
conditions should support the computations, as in \cite{G225}, of the
associativity and braiding matrices for Fibonacci anyons.

For practical purposes, we stay close to the coherence conditions
already available in the literature for structures resembling some of
our rig witnesses. Let us briefly summarize the relevant literature.

If we consider either addition by itself or multiplication by itself,
then the definition of rigs requires that we have a commutative
monoid. An analogous structure has been studied in category theory,
namely symmetric monoidal categories, and suitable coherence conditions
were found by  Mac Lane \cite{maclane1} and simplified by Kelly
\cite{kelly}.  Here is an example to clarify what ``suitable'' means.

\begin{ex} \label{coh-assoc} 
  In ordinary (not witness) algebra, the associative law for addition,
  $(A+B)+C=A+(B+C)$, 
  implies that one can safely omit parentheses in sums of any number
  of terms.  For example, one can deduce $((A+B)+C)+D=A+(B+(C+D))$,
  and similarly for more summands and for other arrangements of the
  parentheses.  In fact, for the specific case of
  $((A+B)+C)+D=A+(B+(C+D))$, two deductions are available. One goes
  via $(A+B)+(C+D)$, and the other goes via $(A+(B+C))+D$ and
  $A+((B+C)+D)$. 

  In witness algebra, the corresponding facts are as follows. Given
  witnesses $\alpha^+_{A,B,C}$ as above, for all raw elements $A,B,C$,
  we can construct, by composing them, witnesses for
  $((A+B)+C)+D=A+(B+(C+D))$ and similarly for more summands and for
  other arrangements of the parentheses.  In fact, for the specific
  case of $((A+B)+C)+D=A+(B+(C+D))$, two such compositions are
  available, corresponding to the two deductions in ordinary algebra.
  A typical coherence condition would require that these two
  compositions coincide. \textqed
\end{ex} 

If we use associativity to rearrange parentheses in sums with more
than four summands, there will, in general, be many deductions for a
single equation in ordinary algebra, and therefore many witnesses,
composites of witnesses of type $\alpha^+$, for the same equation in
witness algebra.  We would like all these witnesses for the same
equation to coincide. So, a priori, we would impose infinitely many
coherence conditions, with more and more variables.  Fortunately, Mac
Lane showed in \cite{maclane1} that the coherence condition for
associativity with four summands implies all the other coherence
conditions for associativity.  Moreover, he found a small number of
coherence conditions for associativity, unit, and commutativity that
imply that, whenever two reasonable compositions of these witnesses
witness the same equation, they coincide.  (``Reasonable'' requires
careful formulation, to avoid, for example, expecting the special case
$\w{\gamma^+_{A,A}}{A+ A}{A+ A}$ of commutativity to coincide with
$1_{A+A}$.)  We shall adopt Mac Lane's coherence conditions, as
simplified by Kelly \cite{kelly}, for the additive structure of our
braid semirings, i.e., for the rig witnesses of the forms
$\alpha^+,\lambda^+,\rho^+,\gamma^+$.

Although it seems natural and simple to treat multiplication the same
way, we shall not adopt these same conditions for the multiplicative
structure.  The reason lies in the behavior of anyons that we intend
to model.  Here is a rough explanation of the situation; a more
detailed (and thus more accurate) explanation can be found in
\cite{G225, PP}.  Think of a product $A\times B$ as representing an
anyon (or anyon system) of type $A$ located next to one of type $B$.
The commutativity witness
$\w{\gamma^\times_{A,B}}{A\times B}{B\times A}$ represents
interchanging the locations of $A$ and $B$.  Anyons inhabit a
two-dimensional space, and so there are two different ways to move $A$
from, say, the left of $B$ to the right of $B$: $A$ could pass in
front of $B$ or behind it. If we (arbitrarily) take
$\gamma^\times_{A,B}$ to represent the transposition that moves $A$ in
front of $B$, then $\w{(\gamma^\times_{B,A})^{-1}}{A+B}{B+A}$
represents the transposition moving $A$ behind $B$, and we do not want
these to always coincide.  In other words, we do not want the
so-called \emph{symmetry condition}
\[
\gamma^\times_{A,B}*\gamma^\times_{B,A}=1_{A+B}
\]
to hold in general. Indeed, the left side of the symmetry condition
represents moving $A$ all the way around $B$,  back to its original
location.  The non-triviality of such \emph{braiding} operations is
the key to the usefulness of anyons in quantum computation.  But the
symmetry condition is among the coherence conditions of Mac Lane and
Kelly.  Our multiplicative structure should therefore be subject only
to some weaker system of coherence conditions, allowing non-trivial
braiding.  

Fortunately, Joyal and Street \cite{js1, js2} have given a system of
coherence conditions that accomplishes exactly what we need.  Their
notion of ``braided monoidal category'' is like ``symmetric monoidal
category'' except that the symmetry condition is omitted (and another
condition, deducible using symmetry but not otherwise, is added).  The
coherence conditions for braided monoidal categories are included
among the axioms for modular tensor categories; see \cite{G225, PP}.  We
shall adopt the Joyal--Street coherence conditions for the
multiplicative structure of our braid semirings. 

Beyond the additive and multiplicative structures, whose coherence
conditions we borrow from Mac Lane, Kelly, Joyal, and Street, we also
have the distributivity witnesses which connect the additive and
multiplicative structures. 

The available literature concerning coherence conditions for
distributivity is the paper \cite{laplaza} of Laplaza. He introduces
and justifies a system of such coherence conditions for the situation
where both the additive and multiplicative structures are symmetric
monoidal structures.  In our situation, however, only the additive
structure is symmetric; the multiplicative structure is merely
braided. As a result, we must modify Laplaza's coherence conditions to
work properly with braided multiplication.  We have carried out this
modification in \cite{cohere} and proved some theorems there that
indicate its appropriateness.  In the present paper, we shall only
record the coherence conditions that we found and some remarks about
them, referring to \cite{cohere} for details.

\begin{conv}    \label{diagrams}
Rather than writing these conditions as equations, we shall exhibit
them as diagrams, in accordance with the following conventions.  Each
diagram will be a directed graph, with vertices labeled by raw
elements and with directed edges labeled by witnesses.  If $\xi$
labels an edge from a vertex $A$ to a vertex $B$ then \w\xi AB.
Consider a path in the underlying undirected graph (obtained by
forgetting the orientations of the edges); note that the edges in such
a path may be directed forward or backward along the path.  Associate
to this path the witness for $A=B$ obtained, by composing (by $*$), in
order along the path, the labels of those edges that are directed
forward along the path and the inverses of the labels of the edges
directed backward along the path.  In our diagrams, the underlying
undirected graph will always be just a cycle, so for each pair of
vertices $A,B$, there will be exactly two paths from $A$ to $B$,
associated with two witnesses for $A=B$. The diagram is to be
interpreted as the equation saying that these two witnesses are equal.
It is easy to check that, if we had chosen two other vertices $A'$ and
$B'$ instead of $A$ and $B$, then the resulting equation between two
witnesses for $A'=B'$ would be equivalent to the equation described
here between witnesses for $A=B$.
(The proof uses clauses~(W4), (W5), and (W6) of the definition of
witness frames.)  

For another equivalent way to interpret the diagram, consider any
vertex $A$ in the cycle and consider a ``path'' that goes from $A$, once
around the cycle (in either direction), ending back at $A$. (We put
``path'' in quotation marks because, strictly speaking, a path should
not have a repeated vertex.)  Associated with this ``path'' is a
witness $\xi$ for $A=A$.  The equations that interpret the diagram as
above are equivalent not only to each other but also to the equation
$\xi=1_A$.
\end{conv}

\subsection{Coherence for Addition and
  Multiplication}   \label{coh-mon} 
Our coherence conditions for the additive structure,
$\{\alpha^+,\lambda^+,\rho^+,\gamma^+\}$, are Kelly's simplification
\cite{kelly} of Mac Lane's coherence conditions \cite{maclane1} for
symmetric monoidal categories.  In the diagram form explained in
Convention~\ref{diagrams}, they are the following Figures~1--4, in
whose captions we have given names for the conditions. (The first
figure, the pentagon, is the previously discussed case of two
witnesses for moving the parentheses, in a sum of four terms, from the
extreme left to the extreme right.)

\begin{figure}[H]
\[\xymatrix@R+1pc{
& (A+ B)+(C+ D)\ar@/^/[rd]^{\ap_{A,B,C+ D}}\\
((A+ B)+ C)+ D\ar@/^/[ru]^{\ap_{A+ B,C,D}}
           \ar[d]_{\ap_{A,B,C}+1_D} 
&& A+(B+(C+ D)) \\
(A+(B+ C))+ D
    \ar[rr]_{\ap_{A,B+ C,D}} 
&& A+((B+ C)+ D) \ar[u]_{1_A+\ap_{B,C,D}}
}\]
\caption{Additive Pentagon Condition}
\end{figure}

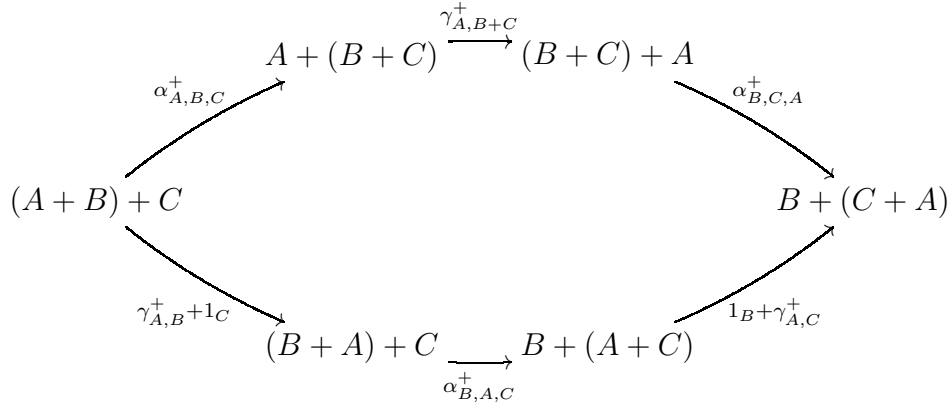
\begin{figure}[H]
\[\xymatrix@C+0pc@R+1pc{
&A+(B+ C)\ar@/^/[r]^{\gp_{A,B+ C}} 
&(B+ C)+ A\ar@/^/[rd]^{\ap_{B,C,A}}\\
(A+ B)+ C\ar@/^/[ru]^{\ap_{A,B,C}}
       \ar@/_/[rd]_{\gp_{A,B}+ 1_C}
&&&B+(C+ A)\\
&(B+ A)+ C\ar@/_/[r]_{\ap_{B,A,C}} 
&B+(A+ C)\ar@/_/[ru]_{1_B+\gp_{A,C}}
}\]
\caption{Additive Hexagon Condition}
\end{figure}

\begin{figure}[H]
\[\xymatrix@C+2pc@R+0pc{
(A+0)+ B\ar@/^/[rr]^{\ap_{A,0,B}} \ar@/_/[rd]_{\rho^+_A + 1_B}
&&A+(0+ B)\ar@/^/[ld]^{1_A+\lambda^+_B}\\
&A+ B
}\]
\caption{Additive Unit Associativity}
\end{figure}
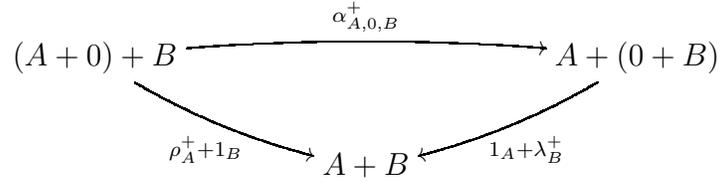

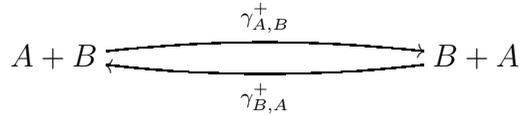
\begin{figure}[H]
\[\xymatrix@C+8pc{
A+ B\ar@/^/[r]^{\gp_{A,B}} &B+ A\ar@/^/[l]^{\gp_{B,A}}
}\]
\caption{Additive Symmetry}
\end{figure}

Our coherence conditions for the multiplicative structure,
$\{\alpha^\times,\lambda^\times,\rho^\times,\gamma^\times\}$, are
those given by Joyal and Street \cite{js1, js2} for braided monoidal
categories, namely the following Figures~5--8. Note that they differ
from the additive ones by the absence of symmetry and the presence of
a second hexagon condition.  This second hexagon condition is like the
first but with every $\gamma^\times_{X,Y}$ replaced with
${\gamma^\times_{Y,X}}$ and the direction of the associated edge
reversed. In view of our Convention~\ref{diagrams} about reading
diagrams as equations, this replacement is equivalent to replacing
every $\gamma^\times_{X,Y}$ with ${\gamma^\times_{Y,X}}^{-1}$. Thus,
the two hexagon conditions are equivalent in the presence of symmetry,
so we needed only one of them in the additive situation. But when
symmetry is unavailable, the two hexagon conditions must both be
assumed.

In the names of the mutiplicative hexagon conditions, ``in front of''
and ``behind'' refer to the way two anyons are interchanged by the
commutativity witnesses $\gamma^\times$.  This corresponds to the
customary picture of braided commutativity in terms of geometric
braids (the same picture that gave the name ``braided'' to this
weakening of symmetry).

\begin{figure}[H]
\[\xymatrix@R+1pc{
& (A\x B)\x (C\x D)
  \ar@/^/[rd]^{\ax_{A,B,C\x D}}\\
((A\x B)\x C)\x D\ar@/^/[ru]^{\ax_{A\x B,C,D}}
           \ar[d]_{\ax_{A,B,C}\x 1_D} 
&& A\x (B\x (C\x  D)) \\
(A\x (B\x C))\x D
    \ar[rr]_{\ax_{A,B\x C,D}} 
&& A\x ((B\x C)\x D) \ar[u]_{1_A\x \ax_{B,C,D}}
}\]
\caption{Multiplicative Pentagon condition}
\end{figure}

\begin{figure}[H]
\[\xymatrix@C+0pc@R+1pc{
&A\x (B\x C)\ar@/^/[r]^{\gx_{A,B\x C}} 
&(B\x C)\x A\ar@/^/[rd]^{\ax_{B,C,A}}\\
(A\x B)\x C\ar@/^/[ru]^{\ax_{A,B,C}}
       \ar@/_/[rd]_{\gx_{A,B}\x  1_C}
&&&B\x (C\x  A)\\
&(B\x A)\x C\ar@/_/[r]_{\ax_{B,A,C}} 
&B\x (A\x C)\ar@/_/[ru]_{1_B\x \gx_{A,C}}
}\]
\caption{Multiplicative Hexagon:
  Moving one factor in front of two}
\end{figure}

\begin{figure}[H]
\[\xymatrix@C+0pc@R+1pc{
&A\x (B\x C) 
&(B\x C)\x A\ar@/_/[l]_{\gx_{B\x C,A}}
  \ar@/^/[rd]^{\ax_{B,C,A}}\\
(A\x B)\x C\ar@/^/[ru]^{\ax_{A,B,C}}
&&&B\x (C\x  A)\ar@/^/[ld]^{1_B\x \gx_{C,A}}\\
&(B\x A)\x C \ar@/^/[lu]^{\gx_{B,A}\x  1_C}
  \ar@/_/[r]_{\ax_{B,A,C}} 
&B\x (A\x C)
}\]
\caption{Multiplicative Hexagon:
  Moving one factor behind two}
\end{figure}

\begin{figure}[H]
\[\xymatrix@C+2pc@R+0pc{
(A\x1)\x B\ar@/^/[rr]^{\ax_{A,1\,B}} 
  \ar@/_/[rd]_{\rho^\x_A \x 1_B}
&&A\x(1\x B)\ar@/^/[ld]^{1_A\x\lambda^\x_B}\\
&A\x B
}\]
\caption{Multiplicative Unit Associativity}
\end{figure}

\begin{ctr}
  The content of this subsection is that our groupoid is equipped with
  two monoidal structures, a symmetric one written with $+$ and a
  braided one written with $\times$. These two structures will be
  connected by distributivity, whose coherence conditions constitute
  the next subsection. 
\end{ctr}

\subsection{Coherence for Distributivity} \label{coh-distrib} 
Our coherence requirements for the distributivity witnesses $\delta$
and $\eps$ are given by Figures~9--18, taken from our paper
\cite{cohere}.  We refer to this paper for motivation and additional
information about these conditions, in particular their connections
with Laplaza's coherence conditions for the case where both $+$ and
$\times$ are symmetric.

\begin{notat}\rm
We shall sometimes use the usual conventions from algebra that $XY$ means
$X\times Y$ and that, for example, $X+ YZ$ means $X+(YZ)$, not
$(X+ Y)Z$. \textqed
\end{notat}

It will be convenient to have the following notation for the deviation
from symmetry.

\begin{notat}
  $\beta_{X,Y}=\gamma^\times_{X,Y}*\gamma^\times_{Y,X}$.
\end{notat}

Thus, symmetry amounts to the requirement that $\beta_{X,Y}=1_{X\times
  Y}$.  In the general braided situation,
\w{\beta_{X,Y}}{X\times Y}{X\times Y}. Pictorially,
if we imagine $\gamma^\times_{X,Y}$ as interchanging $X$ with $Y$ by
moving $X$ in front of $Y$, then $\beta_{X,Y}$ moves $X$ all the way
around $Y$ back to its initial position, first passing in front of $Y$
and then returning behind $Y$.

We now present our coherence conditions for distributivity, along with
some remarks intended to make them easier to understand.

\begin{rmk}             \label{right-distrib}
  We have required witnesses for the distributive law
  $A\times(B+C)=(AB)+(AC)$ but not for the analogous law $(B+C)\times
  A=(BA)+(CA)$.  This is reasonable, since we have commutativity and
  can therefore deduce either of these distributive laws from the
  other.  In terms of witnesses, we have 
\[
\w{\gamma^\times_{B+C,A}*\delta_{A,B,C}*
  (\gamma^\times_{B,A}+\gamma^\times_{C,A})^{-1}} {(B+C)A}{(BA)+(CA)}.
\]
In fact, we have a second, equally good witness for the same equation:
\[
\w{(\gamma^\times_{A,B+C})^{-1}*\delta_{A,B,C}*
(\gamma^\times_{A,B}+\gamma^\times_{A,C})}{(B+C)A}{(BA)+(CA)}.
\]
In terms of the braiding picture of products, the first of these
witnesses moves $A$ behind $B$, $C$, and $B+C$, and the second witness
moves $A$ in front of these other factors. There is no reason to
prefer one of these witnesses to the other, so we shall impose a
coherence condition saying that they are equal. Rather than writing
out that equality, we simplify it a bit by ``clearing fractions'',
i.e., by multiplying both sides by factors to cancel the inverses
that occur in our two witnesses. Once that is done, each side of the
desired coherence condition involves a composition of two
$\gamma^\times$ witnesses, a composition that fits our definition of
$\beta$ above.  Thus, the desired coherence condition takes the simple
form of the left diagram in Figure~9. \textqed
\end{rmk}

\begin{figure}[H]
\begin{minipage}{0.75\textwidth}
\[\xymatrix@C+4pc@R+0pc{
A\x(B+ C)\ar[r]^{\delta_{A,B,C}} 
            \ar[d]^{\beta_{A,B+ C}}
&(A\x B)+(A\x C)\ar[d]^{\beta_{A,B}+\beta_{A,C}}\\
A\x(B+ C)\ar[r]^{\delta_{A,B,C}}
&(A\x B)+(A\x C)}\]
\end{minipage}
\begin{minipage}{0.20\textwidth}
\[
\begin{tikzcd}
   A\x0\ar[out=130, in=50, loop, distance=6em]{}{\beta_{A,0}} & 
\end{tikzcd}
\]
\end{minipage}
\caption{Right Distributive}
\end{figure}

\begin{rmk}
  The right diagram in Figure~9 is essentially the analog of the left
  for a sum of no summands in place of the sum $B+C$ of two
  summands. The precise analog would result from the left diagram by
  changing the vertex labels on the left to $A\times0$ and on the
  right to $0$, changing the horizontal arrows to $\eps_A$, changing
  the left vertical arrow $\beta_{A,0}$, and changing the right
  vertical arrow to $1_0$. (To see that this last $1_0$ is correct,
  use Proposition~\ref{elem} with $f$ being the 0-ary operation 0.)
  Multiplying by the inverse of $\eps_A$, we simplify the desired
  equality to $\beta_{A,0}=1_{A\times0}$, which is depicted on the
  right side of Figure~9. \textqed
\end{rmk}

\begin{rmk}
  It is worthwhile to keep in mind the two witnesses (equal by the
  left part of Figure~9) described above for $(B+C)A=(BA)+(CA)$.  They
  will occur twice in Figure~17, and that rather large figure becomes
  easier to understand if one realizes that, in the two places
  indicated by dashed lines\footnote{Red in the pdf version of the
    paper}, what looks like a composition of three witnesses can be
  understood as just a witness for distributivity from the right
  rather than the left.
  
  Notice also that we have an analogous pair of witnesses (equal by the
  right part of Figure~9) with no summands rather than two,
\[
\w{\gamma^\times_{0,A}*\eps_A}{0A}0\qquad
\w{(\gamma^\times_{A,0})^{-1}*\eps_A}{0A}0.
\]
\end{rmk}

\begin{rmk}
The next three coherence conditions, Figures~10 through 12, say that
distribution respects additive manipulations --- commutativity,
associativity, and unit properties. That is, given $A\times S$ where
$S$ is a sum, it doesn't matter whether we perform additive
manipulations within $S$ and then apply distributivity or first apply
distributivity and then perform the corresponding manipulations on the
resulting sum. 

In these and subsequent figures, we indicate in the caption the
corresponding condition in \cite{laplaza}.  \textqed
\end{rmk}

\begin{figure}[H]
\[\xymatrix@C+8pc@R+1pc{
A(B+ C)\ar[r]^{\delta_{A,B,C}}
      \ar[d]_{1_A\x\gp_{B,C}}      
&(AB)+(AC)\\
A(C+ B)\ar[r]^{\delta_{A,C,B}}
&(AC)+(AB)\ar[u]_{\gp_{AC,AB}}
}\]
\caption{Distribution Respects Additive Commutativity (Laplaza Cond.~I)}
\end{figure}

\begin{figure}[H]
\[\xymatrix@C+2pc@R+1pc{
A(B+(C+ D))\ar@/^/[r]^{\delta_{A,B,C+ D}}
&AB+ A(C+ D)\ar@/^/[r]^{1_{AB}+\delta_{A,C,D}}
&AB+(AC+ AD)\\
A((B+ C)+ D) \ar@/_/[r]_{\delta_{A,B+ C,D}}
           \ar[u]^{{1_A}\x\ap_{B,C,D}}
& A(B+ C)+ AD\ar@/_/[r]_{\delta_{A,B,C}+1_{AD}}
&(AB+ AC)+ AD\ar[u]_{\ap_{AB,AC,AD}}}
\]
\caption{Distribution Respects Additive Associativity (Laplaza Cond.~V)}
\end{figure}

\begin{figure}[H]
\[\xymatrix@C+8pc@R+1pc{
A(B+0)     \ar[r]^{\delta_{A,B,0}}
           \ar[d]^{1_A+\rho^+_B}
&(AB)+(A0) \ar[d]^{1_{AB}+\eps_A}\\
AB
&(AB)+0     \ar[l]^{\rho^+_{AB}}
}\]
\caption{Distribution Respects 0 as neutral (Laplaza Cond.~XXI)}
\end{figure}

\begin{rmk}
  Next are four coherence conditions saying that, when distributing a
  product of several factors across a sum, it doesn't matter whether
  one distributes the whole product at once or the individual factors
  one after the other.  The case of a product of two factors
  distributing across a sum of two summands is the obvious one; it
  implies (in the presence of the other coherence conditions) the
  cases with more factors or summands. It is, however, also necessary
  to cover the cases where the number of factors or the number of
  summands is zero. So we get the four coherence conditions in
  Figures~13 through 16. In our names for the conditions, the numbers
  2 or 0 refer first to the number of factors and second to the number
  of summands. \textqed
\end{rmk}

\begin{figure}[H]
\[\xymatrix@C+6pc@R+1pc{
(AB)(C+ D)\ar[r]^{\ax_{A,B,C+ D}}
&A(B(C+ D))\ar[d]^{1_A\x\delta_{B,C,D}}\\
&A((BC)+(BD))\ar[d]^{\delta_{A,BC,BD}}\\
((AB)C)+((AB)D) \ar@{<-}[uu]^{\delta_{AB,C,D}}
&(A(BC))+(A(BD)\ar@{<-}[l]^{\ax_{A,B,C}+\ax_{A,B,D}}
}\]
\caption{Sequential Distribution $2\x2$ (Laplaza Cond.~VI)}
\end{figure}

\begin{figure}[H]
\[\xymatrix@C+8pc@R+1pc{
(AB)0  \ar[r]^{\ax_{A,B,0}}
       \ar[d]_{\eps_{AB}}
&A(B0) \ar[d]^{1_A\x\eps_B}\\
0
&A0    \ar[l]^{\eps_A}
}\]
\caption{Sequential Distribution $2\x0$ (Laplaza Cond.~XVIII)}
\end{figure}

\begin{figure}[H]
\[\xymatrix@C+2pc@R+1pc{
1(A+ B)      \ar@/^/[rr]^{\delta_{1,A,B}} 
            \ar@/_/[rd]_{\lambda^\x_{A+ B}}
&&(1A)+(1B) \ar@/^/[ld]^{\lambda^\x_A + \lambda^\x_B}\\
&A+ B
}\]
\caption{Sequential Distribution $0\x2$ (Laplaza Cond.~XXIII)}
\end{figure}

\begin{figure}[H]
\[\xymatrix@C+9pc{
1\x0 \ar@/^/[r]^{\eps_1}
     \ar@/_/[r]_{\lambda^\x_0}
&0
}\]
\caption{Sequential Distribution $0\x0$ (Laplaza Cond.~XIV)}
\end{figure}

\begin{rmk}
The remaining coherence conditions for distributivity, in Figures~17
and 18, concern a product of two sums, like
$(A+ B)(C+ D)$. Distributivity lets us expand this as a sum of four
products, but there is a choice whether to apply distributivity first
from the left, obtaining $((A+ B)C)+((A+ B)D)$, or from the right,
obtaining $(A(C+ D))+(B(C+ D))$. One coherence condition (Figure~17)
says that both choices produce the same final result, up to
associativity and commutativity of addition. (Unfortunately, the
associativity and commutativity make the diagram rather large.  It
gets even larger because a single witness for distributivity from the
right looks like a witness for distributivity from the left flanked by
two commutativity witnesses.)  In addition, there are
analogous but far simpler coherence conditions for the case where one
or both of the factors is the sum of no terms rather than of two.  Our
labels for these conditions include numbers 2 or 0 indicating the
number of summands in each factor. \textqed
\end{rmk}

\begin{figure}[H]\footnotesize
\[\xymatrix@C-3pc@R+0pc{
&(A+ B)(C+ D)\ar@/^/[rd]^{\delta_{A+ B,C,D}}
             \ar@/_/[ld]_{\gx_{A+ B,C+ D}}
             \ar@[red]@/^2pc/@{-->}[lddd]\\
(C+ D)(A+ B) \ar[d]_{\delta_{C+ D,A,B}}
&&((A+ B)C)+((A+ B)D)\ar[d]^{\gx_{A+ B,C}+\gx_{A+ B,D}}
                     \ar@[red]@/_9pc/@{-->}[ddd]\\
((C+ D)A)+((C+ D)B)\ar@{<-}[d]_{\gx_{A,C+ D}+\gx_{B,C+ D}}
&&(C(A+ B))+(D(A+ B))\ar[d]^{\delta_{C,A,B}+\delta_{D,A,B}}\\
(A(C+ D))+(B(C+ D))\ar[d]_{\delta_{A,C,D}+\delta_{B,C,D}}
&&((CA)+(CB))+((DA)+(DB)) \ar@{<-}[d]_%
  {(\gx_{A,C}+\gx_{B,C})}^{+\ (\gx_{A,D}+\gx_{B,D})}\\
((AC)+(AD))+((BC)+(BD))\ar@{<-}[d]_{\ap_{AC+ AD,BC,BD}}
&&((AC)+(BC))+((AD)+ BD))\\
(((AC)+(AD))+(BC))+(BD)\ar@{<-}[d]_{\ap_{AC,AD,BC}+1_{BD}}
&&(((AC)+(BC))+(AD))+(BD)\ar[u]_{\ap_{AC+ BC,AD,BD}}
                        \ar[d]^{\ap_{AC,BC,AD}+1_{BD}}\\
((AC)+((AD)+(BC)))+(BD)
&&((AC)+((BC)+(AD)))+(BD)
  \ar@/^3pc/[ll]^{(1_{AC}+\gp_{BC,AD})+1_{BD}}
}\]
\caption{Expand $2\x2$ (Laplaza Cond.~IX)}
\end{figure}

\begin{figure}[H]
\begin{minipage}{0.75\textwidth}
\[\xymatrix@C+2pc@R+1pc{
(A+ B)0\ar[r]^{\gx_{A+ B,0}}
&0(A+ B)\ar[r]^{\delta_{0,A,B}}
&(0A)+(0B)\ar@{<-}[d]^{\gx_{A,0}+\gx_{B,0}}\\
0\ar@{<-}[u]^{\eps_{A+ B}}
&0+0\ar[l]^{\lambda^+_0}
&(A0)+(B0)\ar[l]^{\eps_A + \eps_B}}\]
\end{minipage}
\begin{minipage}{0.18\textwidth}
\[
\begin{tikzcd}
   0\x0\ar[out=130, in=50, loop, distance=6em]{}{\gx_{0,0}} & 
\end{tikzcd}
\]
\end{minipage}
\caption{Expand $2\x0$ and $0\x0$ (Laplaza Conds.~XII and~X)}
\end{figure}

This completes our list of coherence conditions and allows us to
define braid rigs.

\begin{df}
A \emph{braid rig} (or \emph{braid semiring}) is a witness algebra for
the signature $\{+,\times,0,1\}$ together with chosen witnesses of
the forms $\alpha^+,\lambda^+,\rho^+,\gamma^+,
\alpha^\times,\lambda^\times,\rho^\times,\gamma^\times,\delta,\eps$
specified above and subject to the coherence conditions in
subsections~3.1, 3.3, and 3.4.
\end{df}

We emphasize that the specifications of the chosen witnesses in braid
rigs make the associated true algebras into rigs.

\section{Unitary Fusion Semirings}           \label{sec:fusion}

In this section, we describe the particular braid rigs that are used
in anyon models. We begin by describing the true rigs, as this
description will be rather straightforward. Afterward, we shall
describe the raw elements and witnesses.

\subsection{True Algebra}
The true rigs associated to our unitary fusion rigs wlll be rigs in
which
\begin{lsnum}
  \item there is a finite set $\{x_0,x_1,\dots,x_q\}$ of additive
    generators, 
\item each element is a finite sum of generators in a unique way (up to
  order and parentheses), and
\item one of the generators is the multiplicative unit element $1$.
\end{lsnum}
The first two requirements in this list say that, as far as the
additive structure of the rig is concerned, it is the free commutative
monoid on the finite set $\{x_0,x_1,\dots,x_q\}$ of generators.  Note
that the finite sums in requirement~(2) include the empty sum
$0$. In connection with requirement~(3), we adopt the convention
that the generators are numbered so that $x_0=1$.

We also adopt the standard convention that $nx$, for a natural number
$n$ and a rig element $x$, means the sum of $n$ copies of $x$.

Apart from $x_0=1$ and our intention to produce a rig, no requirements
are imposed here on the multiplicative structure.  Of course, the
distributive law for rigs and requirement~(2)  together imply that the
multiplicative structure is completely determined by the products of
the generators. Thus, in the true rig associated to any unitary fusion
rig, we shall have a system of equations of the form 
\[
x_ix_j=\sum_kN^k_{ij}x_k,
\]
where the $N^k_{ij}$ are natural numbers.  These equations, which in
anyon theory are usually called \emph{fusion rules}, suffice to
determine the whole true rig.  They define the products of the
generators, and we extend the definition to arbitrary elements, i.e.,
sums of generators, by distributivity.

The coefficients $N^k_{ij}$ in the fusion rules, called \emph{fusion
  coefficients}, are subject to several constraints, because the
multiplication operation is required to be associative and commutative
with $x_0$ as the unit element. Specifically, to ensure that
$x_0x_i=x_ix_0=x_i$, we must have 
\begin{eqnum}   \label{id}
  N^k_{i0}=N^k_{0i}=\delta_{ik}\qquad\text{for all }i,k,
\end{eqnum}%
where $\delta$ is the Kronecker delta. To ensure that $x_ix_j=x_jx_i$,
we must have
\begin{eqnum}   \label{comm}
  N^k_{ij}=N^k_{ji}\qquad\text{for all }i,j,k.
\end{eqnum}%
To ensure the associativity equation $(x_ix_j)x_k=x_i(x_jx_k)$, we
must have 
\begin{eqnum}   \label{assoc}
  \sum_rN^r_{ij}N^l_{rk}=\sum_sN^s_{jk}N^l_{is}\qquad\text{for all }i,j,k,l.
\end{eqnum}%
(The two sides of this equation are simply the coefficients of $x_l$
in the two sides of the associativity equation.) It is well known that
the associativity, commutativity, and unit laws, which we have ensured
for the generators by means of these constraints on the fusion
coefficients, imply the corresponding laws for all elements, because
products of arbitrary elements were defined from the products of
generators via distributivity.

To summarize this description of the true rigs associated to unitary
fusion rigs: Such a true rig is completely specified by a positive
integer $q$ and a system of fusion coefficients subject to the
constraints \eqref{id}, \eqref{comm}, and \eqref{assoc}. The additive
structure is a free monoid on $\{x_0(=1),x_1,\dots,x_q\}$, and the
multiplicative structure is given by the fusion rules and
distributivity. 

\subsection{Witness Frame}
We now turn from the true rigs to the full structure of unitary fusion
rigs, i.e., to the raw elements and witnesses. Of course, these must
be defined in a way that produces true algebras of the sort described
above. 

For the rest of this section, we work with a fixed set of fusion
rules, and we use the notation $N^k_{ij}$ as above for the fusion
coefficients.  In particular, we have a fixed value for $q$, the number
of generators other than $1$.

The raw elements are, by definition, all of the $(q+1)$-tuples of finite-dimensional complex Hilbert spaces
\[
\h A=(A_0,A_1,\dots,A_q),
\]
each with a specified orthonormal basis, such that the elements of
each $A_i$ are the formal linear combinations (over \bbb C)
of basis elements.  When we speak of basis
elements, we always mean elements of the specified bases.

The witnesses $\xi$ for the equality of two raw elements,
$\h\xi\vdash\h A=\h B$, are all of the $(q+1)$-tuples of Hilbert-space
isomorphisms (i.e., unitary transformations) between the corresponding
components of $\h A$ and $\h B$,
\[
\h\xi=(\xi_0,\xi_1,\dots,\xi_q) \qquad\text{where }\xi_i:A_i\to B_i.
\]
It is \textbf{not} required that the unitary transformations $\xi_i$
respect the specified bases.  When they do, i.e., when each $\xi_i$
is induced by a bijection between the specified bases of $A_i$ and of
$B_i$, we call $\h\xi$ a basic witness.

The reflexivity witnesses are just $(q+1)$-tuples of identity
maps. Composition and inversion of witnesses are done componentwise. 


\subsection{Addition}

Raw elements are added by forming the componentwise direct sum of the
Hilbert spaces. In more detail, the $i\th$ component of $\h A+\h
B$ has as its specified basis the disjoint union of the specified
bases of the components $A_i$ and $B_i$ of \h A and \h B,
respectively. 

Before proceeding further, we need to add some details about the
additive structure just defined. There are several ways to formally
define direct sums of vector spaces.  The different ways produce
isomorphic results, so it usually doesn't matter which way one
chooses. In our situation, though, our witnesses are themselves
(tuples of) isomorphisms, and when we need to manipulate these
witnesses, it will not do to say that things are well-defined up to
isomorphism.  We therefore describe a few ways to formalize direct
sums and, afterward, indicate a notation that will be convenient for
the rest of our work.

(1) The most common construction of the direct sum $A\oplus B$ of two
vector spaces $A$ and $B$ is the set of ordered pairs $(a,b)$ with
$a\in A$ and $b\in B$.  Addition and scalar multiplication are defined
componentwise.  

Theoretically, this works well, but it becomes awkward in some
situations that we shall have to consider.  Notice that, in direct
sums of three vector spaces we shall have elements of the form
$((a,b),c)$ in $(A\oplus B)\oplus C$ and elements of the form
$(a,(b,c))$ in $A\oplus(B\oplus C)$. Mathematicians frequently ignore
the distinction, because of the obvious isomorphism, but in our
situation, the obvious isomorphism $\alpha^+_{A,B,C}$ is part of
the structure we are defining, so it cannot simply be swept under the
rug. Of course, with more than three summands, we would have an even
greater proliferation of parenthesis patterns, making it more
difficult to see the structures involved. 

One could also introduce sums of the form $A\oplus B\oplus C$ (without
parentheses) as a vector space of ordered triples $(a,b,c)$, and,
depending on one's set-theoretic conventions, such a triple might or
might not be considered the same as one (but not both) of $((a,b),c)$
and $(a,(b,c))$. Similarly for direct sums of more vector spaces.
When we consider multiplication of our raw elements, we shall need to
deal with direct sums of many vector spaces at a time, with no natural
parenthesization of the summands, nor even a really natural ordering.
Representing elements of the direct sum by tuples (or by tuples of
tuples of \dots) becomes increasingly arbitrary and awkward.

(2) Another way to construct direct sums like $A\oplus B$ is to begin
with the specified bases for $A$ and for $B$ and to take the disjoint
union of these bases as a basis for $A\oplus B$; the other elements of
$A\oplus B$ are then formal linear combinations of these basis
elements.

An immediate difficulty here concerns the notion of disjoint union:
What if the two bases are not disjoint? (As an extreme example, we
might have $A=B$ with the same specified basis.) Fortunately, there is
an easy solution, namely to tag the elements of our bases, so that the
disjoint union consists of elements $(a,0)$ and $(b,1)$ with $a$ and
$b$ in the bases of $A$ and of $B$, respectively. The tag notation can
be extended to non-basis vectors from $A$ and $B$. Given a vector
$x\in A$, the corresponding vector in $A\oplus B$ is obtained by
expanding $x$ as a linear combination of basis vectors and replacing
each of those basis vectors $a$ by $(a,0)$.  We call the resulting
vector $(x,0)$. Similarly, if $y\in B$, the corresponding vector in
$A\oplus B$ is called $(y,1)$.

This approach works well for a direct sum of many spaces, even if
these are given as an indexed family without any particular
ordering. We can take the basis elements of $\bigoplus_{i\in I}A_i$ to
be tagged elements of the specified bases of the $A_i$'s, i.e., ordered
pairs $(a,i)$ with $i\in I$ and $a\in A_i$.  This observation will be
useful because such naturally indexed but not naturally ordered direct
sums will occur in our discussion of multiplication of raw elements.
Incidentally, note that, if we imposed some arbitrary ordering on the
indices and then used the traditional approach (1) above, then what we
have written as $(a,i)$ here would be an $|I|$-tuple with one
component equal to $a$ and all the other components equal to 0; the
location of the $a$ would encode $i$ (in effect, $i$ is written in
unary notation).

(3) There is a way to attach tags to vectors, as in (2), without
the need for specified bases.  In (2), we did this as syntactic
sugar, writing $(x,t)$ , when $x$ is a (possibly non-basis) vector and
$t$ is a tag, for something constructed out of tagged basis vectors.
Without resorting to syntactic sugar, we can achieve the same goal as
follows. Select some 1-dimensional Hibert spaces (copies of $\bbb C$),
one space $\bbb C_t$ for each tag $t$ that we might want to use, and
let \ket t be a fixed unit vector in $\bbb C_t$ (the copy in that
space of $1\in\bbb C$). Then define the direct sum $A\oplus B$ to
consist of formal sums of vectors from $A\otimes C_0$ and
$B\otimes C_1$. In general, $\bigoplus_{i\in I}A_i$ consists of formal
sums of vectors from the spaces $A_i\otimes\bbb C_i$.  Because
$\bbb C_i$ is one-dimensional and spanned by \ket i, every vector in
$A_i\otimes\bbb C_i$ has the form $a\otimes\ket i$ for a unique
$a\in A_i$.  A fairly common simplification of the notation for tensor
products would write $a\otimes\ket i$ as just $a\ket i$, which brings
us back to almost the same notation as in (2).

(4) Having introduced witness algebra as a generalization of universal
algebra, we mention another viewpoint that we hope will appeal to
universal algebraists. In each of the preceding three approaches, a
vector $a$ from a summand $A_j$ appears in $\bigoplus_{i\in I}A_i$ as $a$
with additional information that indicates the value of $j$. In (1),
the additional information is the location of the component $a$ amid
many other components (and parentheses) in a tuple (of tuples \dots);
in (2) it is the tag $j$ in $(a,j)$; and in (3) it is the tag $j$ in
$a\ket j$. We can view any such arrangement of tags as an operation
(in the sense of universal algebra) applied to $a$.  Nesting of tags
becomes composition of operations.  This more abstract point of view
allows considerably more freedom in tagging.  We have not yet had need
for this freedom, but it may prove useful in further studies.

\begin{conv}    \label{tag-conv}
For the purposes of this paper, we shall use the tag notation as in
(2), including the use of tags with non-basis vectors. It will do no
harm if the reader views $(a,t)$ as syntactic sugar for the $a\ket t$
of (3), nor will it do harm if the reader views the tagging operation
$(-,t)$ as in (4).
\end{conv}

In view of the definition of the addition operation on raw witnesses,
it is clear that the raw element $0=(0,0,\dots,0)$ consisting of
zero-dimensional Hilbert spaces serves as the additive identity
element. 

Addition of witnesses is defined componentwise in the obvious way.
That is, if $\h\xi\vdash\h A=\h A'$ and $\h\eta\vdash\h
B=\h B'$, then $\h\xi+\h\eta$ has as its $i\th$ component the
isomorphism $A_i+B_i\to A'_i+B'_i$ given by
\[
(a,0)\mapsto(\xi_i(a),0)\qquad\text{and}\qquad
(b,1)\mapsto(\eta_i(b),1).
\]

This completes our description of how addition works in our unitary
fusion rig.  Notice that this structure depends on our fixed fusion
rules only through $q$, the number of non-1 generators. The fusion
coefficients $N_{i,j}^k$ will affect only the multiplicative structure.

Notice also the following property of witness addition, which will be
generalized later and will be useful in the verification of several of
the requirements for witness algebras.

\medskip 

\textbf {Tag Invariance} (preliminary form): When the sum of witnesses
acts on a vector, it leaves the tags unchanged and merely applies the
summand witnesses in the unique reasonable way.

\medskip

Let us check that addition as defined here for raw elements produces
the desired additive structure in the true algebra.  Two raw elements
are equal in the true algebra if and only if they are componentwise
isomorphic, which means just that corresponding components have the
same dimension.  The elements of the true algebra thus correspond
bijectively to $(q+1)$-tuples of dimensions, natural numbers
$(n_0,n_1,\dots,n_q)$, and thus to the formal sums
\[
\sum_{i=0}^qn_ix_i=n_01+\sum_{i=1}^qn_ix_i
\]
that we want as elements of the true algebra.  This correspondence can
be formalized by defining, for each $i$, the raw element $\h x_i$ to
have \bbb C (with specified basis $\{1\}$) in component $i$ and zero
in all other components.  Then every raw element is equal (i.e.,
componentwise isomorphic) to a unique sum of these $\h
x_i$'s. The equivalence classes in the true algebra of the $\h
x_i$'s serve as the additive generators of the true algebra, and
addition of raw elements corresponds to the formal addition used in
our earlier description of the true rig.

\begin{rmk}     \label{preserve-basic} 
  Addition essentially works on (specified) bases; the Hilbert space
  structure (linear structure and inner product) just comes along for
  the ride.  By this we mean two things. First, the specified bases in
  $\h A+\h B$ are built purely set-theoretically (no linear
  combinations 
  involved) from the specified bases in \h A and \h B.  Second, if
  $\h\xi$ and $\h\eta$ happen to be basic witnesses (recall that
  this means they respect specified bases), then $\h\xi+\h\eta$ is
  also basic.  Everything we have done so far would continue to work
  if raw elements were $(q+1)$-tuples of finite sets rather than
  Hilbert spaces.  
\end{rmk}

\subsection{Multiplication}
The product $\h A\times\h B$ of raw elements
$\h A=(A_0,A_1,\dots,A_q)$ and $\h B=(B_0,B_1,\dots,B_q)$ has in
its $k\th$ component the direct sum of $N^k_{ij}$ copies of the tensor
product $A_i\otimes B_j$ for all $i$ and $j$.  

Quite generally, when forming the tensor product of two Hilbert spaces
$A$ and $B$ with specified bases, we let the specified basis of
$A\otimes B$ consist of ordered pairs $(a,b)$ where $a$ and $b$ range
over the specified bases of $A$ and of $B$, respectively.

In our present situation, when forming a direct sum of many such
tensor products, we must adjoin tags to identify the various
components of the sum.  In accordance with Convention~\ref{tag-conv},
we write a typical basis element of the $k\th$ component of
$\h A\times\h B$ in the form
\[
(a_i,b_j,i,j,t)
\]
where $a_i$ and $b_j$ are basis elements from the components $A_i$ of
\h A and $B_j$ of \h B, and where $1\leq t\leq N^k_{ij}$.  Here
the tags $i$ and $j$ serve to identify the tensor product
$A_i\otimes B_j$ in which $(a_i,b_j)$ is a basis element, and the last
tag $t$ serves to tell which of the $N^k_{ij}$ copies of this tensor
product our basis element is in.  It will often be convenient in
calculations (though not technically required) to add a 
subscript indicating which component of $\h A\times\h B$ this
basis element is in; thus, instead of $(a_i,b_j,i,j,t)$, we may write
$(a_i,b_j,i,j,t)_k$.

Recall that we defined $\h x_i$ to be the raw element consisting of
$\bbb C$ in component $i$ and $0$ in all $q$ other components. Let us
calculate the product of two of these raw elements, say
$\h x_r\times\h x_s$. For any $k$, the $k\th$ component of this
product, as described above, is the direct sum of numerous tensor
products, but, because of the many 0 components in $\h x_r$ and
$\h x_s$, many of these tensor products will be 0. Indeed, the only
non-zero summands $(x_r)_i\otimes(x_s)_j$ occur when $i=r$ and $j=s$,
and those summands are $\bbb C\otimes\bbb C\cong\bbb C$. So the direct
sum of these tensor products will be the direct sum of $N^k_{rs}$
one-dimensional spaces.  Since this happens for every $k$, we see that
the product $\h x_r\times\h x_s$ has the same dimensions, in all
components, as $\sum_k N^k_{rs}\h x_k$.  It follows that the true
elements $x_k$ represented by the raw elements $\h x_k$ satisfy the
given fusion rules.  Thus, our definition of mutiplication of raw
elements produces the correct true algebra.

In particular, we can define the element 1 of our unitary fusion rig
to be $\h x_0$, and this serves as a multiplicative identity element
in the true algebra.

We define the multiplication of witnesses so that the principle of Tag
Invariance applies to them, just as for addition. That is, if
$\h\xi\vdash\h A=\h A'$ and $\h\eta\vdash\h B=\h B'$ then
$\h\xi\times\h\eta$ is the $(q+1)$-tuple whose $k\th$ component
sends $(a_i, b_j,i,j,t)$ to
\[
(\xi_i(a_i),\eta_j(b_j),i,j,t).
\]
We repeat the principle of Tag Invariance, now in its final form,
including both addition and multiplication.

\medskip

\textbf {Tag Invariance:} When the sum or product of witnesses acts on
a vector, it leaves the tags unchanged and merely applies the summand
and factor witnesses in the unique reasonable way.

\medskip

This completes the definition of a witness algebra. To make it into a
braid semiring, we must specify the associativity, commutativity, and unit
witnesses for both addition and multiplication; specify the
distributivity witnesses; verify that these rig witnesses respect
witnessed equalities (Section~\ref{nat}); and verify the coherence
conditions (Sections~\ref{coh-mon} and \ref{coh-distrib}).  

\subsection{Additive Rig Witnesses}
In this subsection, we specify witnesses for the associative,
commutative, and identity laws of addition.

For the associative law,
$(\h A+\h B)+\h C=\h A+(\h B+\h C)$, let us consider what
happens in one component, say the $k\th$, and let us omit, for
brevity, the subscripts $k$.  

We begin by observing that the standard basis vectors for $(A+B)+C$
have three possible forms. Basis elements from $A+B$ look like $(a,0)$
or $(b,1)$, and they provide basis elements $((a,0),0)$ and
$((b,1),0)$ in $(A+B)+C$. In addition, $C$ provides basis vectors
$(c,1)$.  Similarly, we see that the standard basis vectors for
$A+(B+C)$ are of the three forms $(a,0)$, $((b,0),1)$, and
$((c,1),1)$.  Now we can define the associativity isomorphism
$\alpha^+:(A+B)+C\to A+(B+C)$ in the obvious way:
\begin{align*}
  ((a,0),0)&\mapsto (a,0)\\
((b,1),0)&\mapsto ((b,0),1)\\
(c,1)&\mapsto((c,1),1).
\end{align*}
Restoring the subscripts that we omitted for brevity earlier, we
should write this $\alpha^+$ as $(\alpha^+_{\h A,\h B,\h
  C})_k$.
It is the $k\th$ component of the associativity witness
$\alpha^+_{\h A,\h B,\h C}$ for
$(\h A+\h B)+\h C= \h A+(\h B+\h C)$.

Before proceeding to other witnesses for addition, we point out an
important property, which will also hold for many --- but not all ---
of the rig witnesses to be introduced later.  It concerns what happens
to tags (the 0's and 1's in our present situation) and the generic
elements of specified bases (the $a,b,c$ in our present situation).

\medskip

\textbf{Tag Manipulation:} All rig witnesses, with the exception of the multiplicative associativity and commutativity witnesses $\alpha^\times$ and $\gamma^\times$, merely manipulate tags, leaving vectors from the given Hilbert spaces unchanged.  These manipulations of the tags do not depend on the particular vectors from the given Hilbert spaces.

\medskip
The exceptional witnesses $\alpha^\times$ and $\gamma^\times$ are what makes
unitary fusion rigs interesting and useful for quantum computation.

The two principles of Tag Invariance (for sums of witnesses) and Tag
Manipulation (for $\alpha^+$) together imply that $\alpha^+$ respects
witnessed equality in the sense explained in Section~\ref{nat}.  That
is, if $\xi\vdash A=A'$, $\eta\vdash B=B'$, and 
$\zeta\vdash C=C'$ then 
\[
  \alpha^+_{A,B,C}*(\xi+(\eta+\zeta))
  =((\xi+\eta)+\zeta)*\alpha^+_{A',B',C'}
\]
Rather than doing a detailed calculation to verify this, we just
notice that manipulations of tags that ignore the input vectors (as in
the $\alpha^+$'s) and manipulations of the input vectors that ignore
the tags (as in both versions of $\xi+\eta+\zeta$) do not interfere
with each other and can therefore be carried out in either order.

The rest of the additive structure is easier than what we have already
done with associativity.  For commutativity, the obvious witness
$\gamma^+_{\h A,\h B}$ for
$\h A+\h B=\h B+\h A$ has in its $k\th$ component, 
\begin{align*}
  (a,0)&\mapsto(a,1)\\(b,1)&\mapsto(b,0).
\end{align*}

The raw zero element $\h0$ is the $(q+1)$-tuple of zero-dimensional
Hilbert spaces.  The required witnesses for $\h A+\h0=\h A$ and
$\h0+\h A=\h A$ are given on basis vectors and therefore, by
linearity, on all vectors, by (note that there are
no basis vectors in \h0)
\[
(a,0)\mapsto a\qquad\text{and}\qquad (a,1)\mapsto a,
\]
respectively.

These witnesses also satisfy the Tag Manipulation principle, and it
follows, just as in the case of associativity, that they respect
witnessed equalities. Furthermore, it is easy to check the coherence
conditions for the additive structure, Figures~1--4 in
Section~\ref{coh-mon}.

\subsection{Multiplicative Rig Witnesses, Part 1}

In this subsection, we handle only the multiplicative identity witnesses.
Associativity and commutativity are more complicated and will be
treated later.  

The raw mutiplicative identity element 1 is the $(q+1)$-tuple with a
one-dimensional space in component 0 and 0-dimensional spaces in all
$q$ of the other components.  In the $0\th$ component, the 
1-dimensional Hilbert space is \bbb C with basis $\{1\}$.

The witness for $1\times\h A=\h A$ has in its $k\th$ component the
isomorphism given on basis elements (and therefore by linearity on all
elements) by
\[
(1, a,0,k,1)\mapsto a.
\]
To see that this makes sense, notice that the $k\th$ component of
$1\times\h A$ has, according to the definition of $\times$, basis
elements of the form $(u, a,i,j,t)$ with $1\leq t\leq N^k_{i,j}$, with
$u$ a basis element of $1_i$, and with $a$ a basis element of
$A_j$. But for all $i\neq0$, we have that $1_i$ is zero-dimensional,
and so it has no basis vectors. So we get elements $(u, a,i,j,t)$ only
when $i=0$, and then $u$ is the number $1\in\bbb C$.  But then
$N^k_{i,j}=N^k_{0,j}=\delta_{j,k}$ is zero for all $j\neq k$, so there
are no values of $t$ available. So we get elements $(u, a,i,j,t)$ only
when $j=k$.  And then $N^k_{0,k}=1$, so the only available value for
$t$ is 1.  So all our basis elements for the $k\th$ component of
$1\times\h A$ are of the form $(1, a,0,k,1)$.  These are in
one-to-one correspondence with the basis elements $a$ of $A_k$ (since
$j=k$), and that provides our witness $\lambda^\times_{\h A}$ for
$1\times\h A=\h A$.

Similarly, we define the required witness $\rho^\times_{\h A}$ for
$\h A\times 1=\h A$ to have, in its $k\th$ component, the
isomorphism
\[
(a,1,k,0,1)\mapsto a.
\]

As before, we still have the Tag Manipulation principle and therefore
these witnesses respect witnessed equality.

\subsection{Distributivity}
 
We must specify witnesses $\delta_{\h A,\h B,\h C}$ for
$\h A\times(\h B+\h C)=(\h A\times\h B)+(\h A\times\h C)$.  Standard
basis elements for components of $B+C$ have two possible forms, namely
$(b,0)$ and $(c,1)$ with $b$ and $c$ in the standard bases for the
corresponding components of $\h B$ and $\h C$. Therefore standard basis
elements for the $k\th$ component of $\h A\times(\h B+\h C)$ have the
possible forms $(a,(b,0),i,j,t)$ and $(a,(c,1),i,j,t)$, with
$1\leq t\leq N^k_{i,j}$. In the $k\th$ component of
$(\h A\times\h B)+(\h A\times\h C)$, we have elements of the two forms
$((a,b,i,j,t),0)$ and $((a,c,i,j,t),1)$. So it is clear how to set up
the desired isomorphism in accordance with the Tag Manipulation
principle: 
\begin{align*}
  (a,(b,0),i,j,t)&\mapsto((a,b,i,j,t),0)\\
(a,(c,1),i,j,t)&\mapsto((a,c,i,j,t),1).
\end{align*}

Similarly, for
$(\h B+\h C)\times\h A=(\h B\times\h A)+(\h C\times\h A)$, we have the
witness in accordance with Tag Manipulation
\begin{align*}
    ((b,0),a,i,j,t)&\mapsto((b,a,i,j,t),0)\\
((c,1),a,i,j,t)&\mapsto((c,a,i,j,t),1).
\end{align*}

Recall that, when we defined braid semirings, only the left
distributivity witnesses were taken as primitive parts of the
structure; the right distributivity witnesses were defined in terms of
the left ones and the commutativity witnesses.  Here, in contrast, we
have specified both left and right distributivity witnesses via Tag
Manipulation.  Thus, when we define the multiplicative commutativity
witnesses $\gamma^\times$, we shall need to ensure that they cohere
with what we have done here, i.e., that the commutativity witnesses
commute appropriately with the left and right distributivity
witnesses. 

With distributivity, we should also include the case where the
addition has no summands, i.e., $\h A\times 0=0$ and $0\times\h A=0$.
In both cases, the raw elements are identical; they have empty bases
in all components.  So the desired witnesses can (indeed must) be
taken to be identity isomorphisms (i.e., reflexivity witnesses).

Because of Tag Manipulation, we obtain, by the same argument as
before, that our distributivity witnesses respect witnessed equality.

\subsection{Coherence}          \label{coherence}
At this point, we have defined all the rig witnesses for a unitary
fusion rig, except for the associativity and commutativity of
multiplication, $\alpha^\times$ and $\gamma^\times$. All the witnesses
defined so far satisfy the Tag Manipulation principle. Because of this
principle  and the Tag Invariance principle for the operations on
witnesses, we know that the rig witnesses defined so far respect
witnessed equality as required in Section~\ref{nat}.

In fact, the Tag Manipulation principle also gives us many of the
coherence conditions required in Sections~\ref{coh-mon} and
\ref{coh-distrib}. Specifically, Figures~1, 2, 3, 4, 10, 11, 12, and
15 involve no $\alpha^\times$ or $\gamma^\times$, so all the rig
witnesses in these figures are already defined, and it is routine to
check that these coherence conditions hold.

The same goes for Figure~17 if we use the dashed arrows for right
distributivity rather than the solid arrows that express right
distributivity in terms of left distributivity and $\gamma^\times$.  So, provided we
ensure that our Tag Manipulation definition of right distributivity
witnesses agrees with the definition via $\gamma^\times$, we shall
have verified Figure~17. 

Several more of the coherence conditions are satisfied simply because
there is only one isomorphism from a zero-dimensional Hilbert space to
itself. Thus, any two witnesses of $0=0$ coincide. This gives us the
coherence conditions in the second part of Figure~9, and in
Figures~14, 16, and (both parts of) 18.

What remains to be checked, after we discuss $\alpha^\times$ and
$\gamma^\times$?  

First, we must make sure that our Tag Manipulation
definition for right distributivity witnesses agrees with what we
obtain from the left distributivity witnesses by conjugation with
commutativity witnesses of either form, as displayed in
Remark~\ref{right-distrib}.  That  will ensure not only that our
definitions for right distributivity witnesses are coherent but also
that the coherence condition in the first part of Figure~9 holds.
Indeed, that part of Figure~9 was exactly the statement that the two
formulas in Remark~\ref{right-distrib} agree.

Second, we must verify the coherence conditions for the multiplicative
structure in Figures~5, 6, 7, and 8  (the Joyal-Street conditions for
braided monoidal structure).

Finally, we must make sure that distributivity and multiplicative
associativity cohere as required in Figure~13.

\subsection{Associativity of Multiplication}

We turn to the problem of defining associativity witnesses 
\[
\w{\alpha^\times_{\h A,\h B,\h C}}
{(\h A\times\h B)\times\h C}{\h A\times(\h B\times\h C)}.
\]
With our usual notational conventions, the left side of the
associativity equation, $(\h A\times\h B)\times\h C$, has, in its
$l\th$ component, basis elements of the form
\[
((a,b,i,j,t)_r,c,r,k,u)
\]
with $i$, $j$, and $k$ ranging from 0 to $q$; with $a$, $b$, and $c$
basis elements of the Hilbert spaces $A_i$, $B_j$, and $C_k$,
respectively; and with $1\leq t\leq N^r_{ij}$ and
$1\leq u\leq N^l_{rk}$..  Similarly, the $l$-component of the right
side, $\h A\times (\h B\times\h C)$ has basis elements of the form
\[
(a,(b,c,j,k,v)_s,i,s,w)
\]
with $i,j,k$ and $a,b,c$ as above, $1\leq v\leq N^s_{jk}$, and
$1\leq w\leq N^l_{is}$.  Notice that both forms involve the same basis
elements $a,b,c$ and the same indices $i,j,k,l$ for the elements and
the final result, but that the configurations of tags are quite
different.  Fortunately, equation~\eqref{assoc} above ensures that,
for any fixed $i,j,k,l$, the number of tag configurations is the same
in both cases.

If we try to define the associativity witness $\alpha^\times_{A,B,C}$
in the spirit of the Tag Manipulation principle, then we should set
up a bijection between the two sorts of basis elements,
$((a,b,i,j,t)_r,c,r,k,u)$ and $(a,(b,c,j,k,v)_s,i,s,w)$, that leaves
$a,b,c$ and therefore also $i,j,k$ and $l$ unchanged, but sets up a
bijection between the possible triples of tags $(r,t,u)$ and
$(s,v,w)$.  As noted above, equation~\eqref{assoc} ensures that a
bijection exists, for each fixed $i,j,k,l$, but it is not evident how
or even whether we can choose such bijections coherently.  It turns
out that such a choice of bijections is not possible in general.  We
devote the next subsection to showing why it is impossible in a
specific, rather simple example.

\subsection{Fibonacci Example and Tag Manipulation}     
\label{fib1}
We consider the fusion rules for the Fibonacci anyon model.  This
model has $q=1$, so there are only two generators, $x_0$ and $x_1$.
(They are often called 1 and $\tau$ respectively, but for the time
being we retain the $x_0,x_1$ notation for consistency with previous
sections.)  The fusion rules say that
\[
x_1\times x_1=x_1+x_0 \qquad\text{(i.e., }\tau^2=\tau+1\text{)}
\]
and, as usual, $x_0$ is the multiplicative identity
\[
x_0\times x_1=x_1\times x_0=x_1\quad\text{and}\quad x_0\times
x_0=x_0. 
\]
Thus, the non-zero fusion coefficients are 
\[
N^1_{11}=N^0_{11}=N^1_{01}=N^1_{10}=N^0_{00}=1.
\]
Since all the fusion coefficients are 0 or 1, we can considerably
simplify the notation in the previous subsection: If $t,u,v,w$ exist
at all, they must be equal to 1, so it is not necessary to mention
them.  And the conditions for their existence are precisely that the
corresponding fusion coefficients must be 1 rather than 0.  

Thus, instead of seeking a bijection between triples of tags
$(r,t,u)\leftrightarrow(s,v,w)$ as above, we seek bijections
$r\leftrightarrow s$ such that appropriate $t,u$ exist for $r$ if and
only if appropriate $v,w$ exist for $s$.  

Now let us consider some specific cases for $i,j,k,l$.

\noindent\textbf{Case 1:}
$i=0$

Recalling equation~\eqref{id}, we find that existence of $t$ requires
$r=j$, and existence of $w$ requires $s=l$.  Furthermore, given these
equations, the existence of $u$ and the existence of $v$ require the
same thing, namely $N^l_{jk}=1$.  (If either of $j,k$ is 0, then $l$
equals the other one; if $j=k=1$ then $l$ can be 0 or 1.)  In any
case, we need to chose a bijection between $\{j\}$ and $\{l\}$;
There's only one bijection between two singletons, so that's what we
choose.

\noindent\textbf{Case 2:} 
$k=0$ 

This is symmetrical to Case~1. Existence of $v$ and $u$ requires $s=j$
and $r=l$, respectively, and then the existence conditions for $w$ and
$t$ give the same requirement $N^l_{ij}=1$. So we need to choose a
bijection between $\{l\}$ and $\{j\}$; there's only one bijection, so
we choose it.

\noindent\textbf{Case 3:}
$i=k=1$ but $j=0$

Then existence of $t$ and of $v$ requires $r=s=1$. No additional
requirements arise from existence of $u$ and $v$; either value of $l$
is possible.  But in either case, we again need to choose a bijection
between $\{1\}$ and $\{1\}$; we choose the only bijection there is.

\noindent\textbf{Case 4:}
$i=j=k=1$ but $l=0$

Existence of $u$ and $v$ requires $r=s=1$. No additional requirements
arise from existence of $t$ and $v$.  We again need to choose a
bijection between $\{1\}$ and $\{1\}$; we choose the only bijection
there is.

\noindent\textbf{Case 5:}
$i=j=k=l=1$

Now both values of $r$ and both values of $s$ are available; no
constraints arise from existence of any of $t,u,v,w$.  So we need to
choose a bijection between $\{0,1\}$ and $\{0,1\}$.  

So we have two reasonable attempts to define associativity witnesses
for multiplication according to the Tag Manipulation principle in the
Fibonacci model, namely to use the identity bijection in Case~5 or to
use the ``switching'' bijection $0\leftrightarrow1$. In the other four
cases, we use the only bijection that there is.

Let us see what happens if we use the identity bijection in
Case~5. And let us look at the simplest non-trivial case of the
pentagon condition, namely the case where
$\h A=\h B=\h C=\h D=(0,\bbb C)=\h x_1$.  So all four factors
have the one-dimensional space \bbb C with standard basis element 1 in
their 1-component, and they have the 0-dimensional space with empty
basis in the 0-component.

Before beginning the computations, let us simplify the notation a bit.
We have written the basis elements for a product $X\times Y$ as
$(x,y,i,j,t)$, where $i$ and $j$ indicate which components of $X$ and
$Y$ the basis elements $x$ and $y$ come from, and $t$ is an index
ranging from 1 to a suitable fusion coefficient $N^k_{ij}$.  In our
situation, we can omit $t$, because the only value it ever has is 1.
We can also omit $i$ and $j$ provided we know, in some other way,
which components of $X$ and $Y$ our $x$ and $y$ come from. If $X,Y$
are any of $\h A,\h B,\h C,\h D$ then we know that basis elements come
from the 1 component, because the bases for the 0-components are
empty.  If $X$ and $Y$ are products of several factors, then $x$ and
$y$ would themselves be compound expressions, and we would know which
components they are in if, as mentioned earlier, we append subscripts
indicating the components.  We adopt, for the present computation,
this convention: Tag compound expressions to indicate which component
they lie in, and then omit $i,j,t$ from the standard notation. Thus,
if $(x,y,i,j,1)$ is a basis element of the $k\th$ component of some
product, we shall write it as $(x,y)_k$.

With this notation, the basis elements for the 1-component of
$((\h A\times\h B)\times\h C)\times\h D$ have three possible forms
\[
(((a,b)_1,c)_0,d)_1\qquad(((a,b)_0,c)_1,d)_1\qquad(((a,b)_1,c)_1,d)_1 
\]
where $a,b,c,d$ are basis elements.  Actually, it is unnecessary to
write $a,b,c,d$ here, since these basis elements are all simply 1. All
the real information is in the subscripts.  Nevertheless we continue
to write $a,b,c,d$ to match previous notation.

Let us trace what happens to these three sorts of basis elements in
$((\h A\times\h B)\times\h C)\times\h D$ along the two paths to $\h
A\times(\h B\times(\h C\times\h D))$ in the multiplicative pentagon
condition, Figure~5. 

We begin by following the longer of the two paths, around the bottom
of the figure.  The first witness on that path, for the equation
$((\h A\times\h B)\times\h C)\times\h D=(\h A\times(\h B\times\h
C))\times\h D$,
leaves $D$ alone and applies $\alpha^\times_{\h A,\h B,\h C}$ to the
other three factors.  For the first of our three possible forms, we
are in Case~4, so we get $((a,(b,c)_1)_0,d)_1$.  The second and third
cases are in Case~5, so our decision to use the identity bijection in
this case leads to $((a,(b,c)_0)_1),d)_1$ for the second form and
$((a,(b,c)_1)_1,d)_1$ for the third.  Summarizing, our three forms
have become, after this first step along the long side of the
pentagon,
\[
((a,(b,c)_1)_0,d)_1\qquad ((a,(b,c)_0)_1),d)_1\qquad ((a,(b,c)_1)_1,d)_1.
\]

The next step along this path is the witness for
$(\h A\times(\h B\times\h C))\times\h D =\h A\times((\h B\times\h
C)\times\h D)$,
namely $\alpha^\times_{A,B\times C,D}$.  This time the second form is in
Case~3, so we use the unique available bijection and obtain
$(a,((b,c)_0,d)_1)_1$.  The first and third forms are in Case~5.
Because we're using the identity bijection in this case, we get
$(a,((b,c)_1,d)_0)_1$ for the first form and $(a,((b,c)_1,d)_1)_1$ for
the third. The summary now reads
\[
(a,((b,c)_1,d)_0)_1\qquad (a,((b,c)_0,d)_1)_1\qquad (a,((b,c)_1,d)_1)_1.
\]

The last step on this path of the pentagon is the witness for
$\h A\times((\h B\times\h C)\times\h D)=\h A\times(\h B\times(\h
C\times\h D))$,
which leaves $\h A$ alone but applies $\alpha^\times_{\h B,\h C,\h D}$
to the rest.  For the first form, we have Case~4 and we obtain
$(a,(b,(c,d)_1)_0)_1$.  The second and third forms are in Case~5, and
our decision to use the identity bijection produces
$(a,(b,(c,d)_0)_1)_1$ for the second form and $(a,(b,(c,d)_1)_1)_1$
for the third.  So the summary, for the entire long path in the
pentagon, is
\[
(a,(b,(c,d)_1)_0)_1\qquad (a,(b,(c,d)_0)_1)_1\qquad (a,(b,(c,d)_1)_1)_1.
\]

This completes the calculation for the long path; we return to the
original three forms and calculate what happens to them along the
short path, around the top of the pentagon, still using the identity
bijection in Case~5.

The first step is the witness for
$((\h A\times\h B)\times\h C)\times\h D=(\h A\times\h B)\times(\h
C\times\h D)$,
namely $\alpha^\times_{\h A\times\h B,\h C,\h D}$.  The second form is
in Case~1, so we get $((a,b)_0,(c,d)_1)_1$. The first and third forms
are in Case~5, so our choice of the identity bijection produces
$((a,b)_1,(c,d)_0)_1$ for the first form and $((a,b)_1,(c,d)_1)_1$ for
the third. The current summary is therefore
\[
((a,b)_1,(c,d)_0)_1\qquad ((a,b)_0,(c,d)_1)_1\qquad ((a,b)_1,(c,d)_1)_1.
\]

The remaining step on the short path, the associativity witness for
$(\h A\times\h B)\times (\h C\times\h D)=\h A\times(\h B\times(\h
C\times\h D)$
is $\alpha^\times_{\h A,\h B,\h C\times\h D}$.  The first form is in
Case~2, so we get $(a,(b,(c,d)_0)_1)_1$. The second and third forms
are in Case~5, so our choice of the identity bijection yields
$(a,(b,(c,d)_1)_0)_1$ for the second form and $(a,(b,(c,d)_1)_1)_1$
for the third. The final summary for the short path is
\[
(a,(b,(c,d)_0)_1)_1\qquad (a,(b,(c,d)_1)_0)_1\qquad (a,(b,(c,d)_1)_1)_1.
\]

This is not the same as the summary for the long path.  The third form
yielded the same result for both ways around the pentagon, but the
results for the first and second forms have been interchanged.  This
means that our choice of the identity bijection in Case~5 cannot be
correct; it fails to satsify the pentagon condition.

Would the alternative choice in Case~5, the switching bijection, fare
better?  We could repeat the entire computation above using the
switching bijection in place of the identity, but there is a more
efficient method.  Instead of repeating the calculation, we merely
keep track of the changes. Every time Case~5 occurred in the preceding
calcuation, we must now switch its two outcomes.  Around the long
path, the bottom of the pentagon, we have three switches, which when
composed just interchange the first and second forms in the final
summary, producing
\[
(a,(b,(c,d)_0)_1)_1\qquad (a,(b,(c,d)_1)_0)_1\qquad
(a,(b,(c,d)_1)_1)_1.
\]
Around the short side, we get two switches, which when composed give
a 3-cycle permutation of the final summary, producing 
\[
(a,(b,(c,d)_1)_0)_1\qquad (a,(b,(c,d)_1)_1)_1\qquad (a,(b,(c,d)_0)_1)_1.
\]
The long and short paths still don't agree; in fact none of the three
forms produce the same results for both paths.  So the ``switching''
choice in Case~5 doesn't work either; it violates the pentagon
condition.

\subsection{Associativity and Commutativity of Multiplication}
The preceding computations show that we cannot insist upon the Tag
Manipulation principle for the associativity witnesses for
multiplication.  Nor can we expect the Tag Manipulation principle to
hold for the commutativity witnesses for multiplication.  Indeed, for
the same Fibonacci example used above, once we compute the
associativity isomorphisms $\alpha^\times$ in
Section~\ref{sec:fibonacci}, we shall find that they and the hexagon
conditions require the commutativity isomorphisms $\gamma^\times$ to
violate the Tag Manipulation principle.

This situation is the source of the complexity of finding
anyon models with prescribed fusion rules. Given the fusion rules, we
have produced all of the structure of a unitary fusion rig except for
$\alpha^\times$ and $\gamma^\times$ straightforwardly, according to
the Tag Invariance and Tag Manipulation principles.

The task of producing suitable $\alpha^\times$ and $\gamma^\times$
looks daunting, first because we must define suitable
$\alpha^\times_{\h A,\h B,\h C}$ and $\gamma^\times_{\h A,\h B}$ for
all of the infinitely many raw elements $\h A,\h B,\h C$ and second
because we must satisfy several coherence conditions. Recall that we
found, in Section~\ref{coherence}, that many coherence conditions will
automatically hold, but several, listed at the end of that subsection,
remain as constraints on $\alpha^\times$ and $\gamma^\times$.  

Fortunately, some of the coherence conditions work in our favor,
reducing the number of instances $\alpha^\times_{\h A,\h B,\h C}$ and
$\gamma^\times_{\h A,\h B}$ that we need in order to determine all the
other instances. Specifically, we can obtain the general witnesses
$\alpha^\times_{\h A,\h B,\h C}$ and $\gamma^\times_{\h A,\h B}$ from
the special case where $\h A$, $\h B$, and $\h C$ are among the $q+1$
elements $\h x_i$, as follows.

Let us write (adopting Laplaza's notation in \cite{laplaza})
$\delta^\#$ for the right distributivity witness 
\[
\w{\delta^\#_{\h A,\h B,\h C}}{(\h B+\h C)\h A}{\h{BA}+\h{CA}}
\]
as defined earlier using the Tag  Manipulation principle.  Then we use 
$\delta^\#$ and the left distributivity witness $\delta$ to reduce
commutativity witnesses with sums as subscripts, by setting 
\begin{align}           \label{gamma-reduce}
\gamma^\times_{\h B+\h C,\h A}&=
\delta^\#_{\h A,\h B,\h C}*
(\gamma^\times_{\h B,\h A}+\gamma^\times_{\h C,\h A})*
(\delta_{\h A,\h B,\h C})^{-1}\\
\gamma^\times_{\h A,\h B+\h C}&=
\delta_{\h A,\h B,\h C}*
(\gamma^\times_{\h A,\h B}+\gamma^\times_{\h A,\h C})*
(\delta^\#_{\h A,\h B,\h C})^{-1}\notag
\end{align}
Because every raw element is equal ($=$, not necessarily $\equiv$) to
a sum of $\h x_i$'s and because witnesses must respect witnessed
equalities, these formulas \eqref{gamma-reduce} allow us to represent
all the infinitely many commutativity witnesses
$\gamma^\times_{\h A,\h B}$ that we need to define in terms of just
finitely many (at most $(q+1)^2$) of them, namely those where the
subscripts are $\h x_i$'s.

There is another useful way to view the equations
\eqref{gamma-reduce}. In the preceding paragraph, we worked on the
basis that both $\delta$ and $\delta^\#$ are given (by Tag
Manipulation), and these equations serve to reduce the task of
defining $\gamma^\times$.  Let us now return to the viewpoint of
Section~\ref{sec:braid}, namely that $\delta^\#$ is not primitive but
rather defined from $\delta$ and $\gamma^\times$. Specifically, we
had, in Remark~\ref{right-distrib}, two formulas providing such
definitions, and the coherence condition in the first part of Figure~9
said precisely that these two definitions agree. Notice now that those
two definitions for $\delta^\#$ are equivalent to equations
\eqref{gamma-reduce}.  This has several pleasant consequences.

First, equations \eqref{gamma-reduce} are not arbitrary, nor are they
produced merely by the desire to simplify our task of defining
$\gamma^\times$.  They are forced upon us if we want to have both a
prescribed $\delta^\#$ (from Tag Manipulation) and the definitions of
$\delta^\#$ in Remark~\ref{right-distrib}.  In effect, we need no
longer worry about the two viewpoints espoused in the two preceding
paragraphs; the two agree. 

Second, since the two equations in \eqref{gamma-reduce} give us both
of the proposed definitions of $\delta^\#$ in
Remark~\ref{right-distrib}, we get that those two definitions agree.
That is, we get that the coherence condition in the first part of
Figure~9 holds.  (Recall that we already had the other part of
Figure~9, because it involves witnesses for $0=0$.)

The task of defining $\gamma^\times$ can be reduced a bit more. When
one of the subscripts is 1 (i.e., $\h x_0$), the commutativity witness
is determined by Proposition~2.1 of \cite{js2}, a consequence of the
braided monoidal coherence conditions (Figures~5--8):
\[
\gamma^\times_{A,1}=\rho^\times_A*(\lambda^\times_A)^{-1}\quad
\text{and}\quad
\gamma^\times_{1,A}=\lambda^\times_A*(\rho^\times_A)^{-1}.
\] 
Thus, to define $\gamma^\times$, it suffices to define
$\gamma^\times_{\h x_i,\h x_j}$ with $i,j=1,2,\dots,q$.

Similar simplifications are possible for the multiplicative
associativity witnesses $\alpha^\times$.  Whenever one of the
subscripts of $\alpha^\times$ is a sum, we can express that witness in
terms of $\alpha^\times$ with the individual summands as subscripts
together with suitable distributivity witnesses $\delta$.  Figure~13
gives this information when the sum is in the third subscript:
\[
\alpha^\times_{A,B,C+D}=\delta_{AB,C,D}*
(\alpha^\times_{A,B,C}+\alpha^\times_{A,B,D})*
(\delta_{A,BC,BD})^{-1}*
(1_A\times\delta_{B,C,D})^{-1}.
\]
Analogous information with the sum in the first or second subscript
was deduced from our coherence conditions in \cite{cohere}.  We
reproduce here Figures~20 and 24 from that paper:

\begin{figure}[H]\small
\[\xymatrix@C+1pc@R+0pc{
(BA)(C+ D)       \ar@/^/[rr]^{\delta_{BA,C,D}}
&& (BA)C+(BA)D  \ar[d]^{\gx_{BA,C}+\gx_{BA,D}}
\\
(C+ D)(BA)       \ar@{<-}[u]^{\gx_{BA,C,D}}
                \ar@{-->}[rr]^{\delta^\#_{C,D,BA}}
&&C(BA)+ D(BA)   \ar@{<-}[d]^{\ax_{C,B,A}+\ax_{D,B,A}}
\\
((C+ D)B)A       \ar[u]^{\ax_{C+ D,B,A}}
                \ar@{-->}[rddd]^{\delta^\#_{C,D,B}\x1_A}
&&(CB)A+(DB)A   \ar@{<-}[d]^{\gx_{A,CB}+\gx_{A,DB}}
\\
(B(C+ D))A       \ar[u]^{\gx_{B,C+ D}\x1_A}
&&A(CB)+ A(DB)   \ar@{<-}[d]^{\delta_{A,CB,DB}}
\\
((BC)+(BD))A    \ar@{<-}[u]^{\delta_{B,C,D}\x1_A}
&&A((CB)+(DB))  \ar@/^/[ld]^-{\gx_{A,CB+ DB}}
\\
&((CB)+(DB))A   \ar@/^/@{<-}[lu]^-{(\gx_{B,C}+\gx_{B,D})\x1_A}
                \ar@{-->}[ruuu]^{\delta^\#_{CB,DB,A}}
}\]
\caption{Laplaza Cond.~VII}
\end{figure}
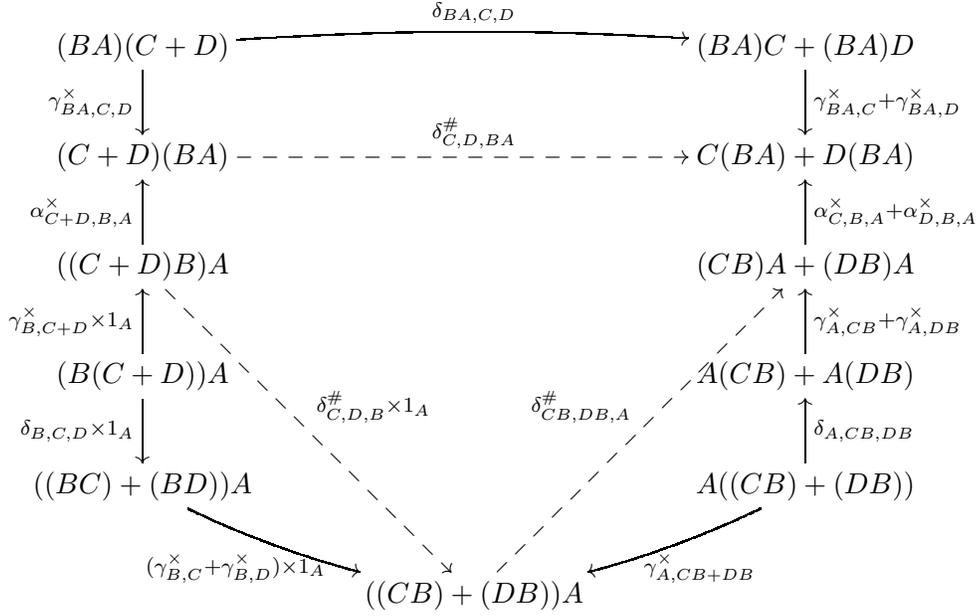

\begin{figure}[H]\small
\[\xymatrix@C+1pc@R+0pc{
A((BC)+(BD))       \ar@/^2pc/[r]^{1_A\x(\gx_{B,C}+\gx_{B,D})}
&A((CB)+(DB))      \ar@/^2pc/[r]^{\delta_{A,CB,DB}}
&A(CB)+ A(DB)      \ar@{<-}[d]^{\ax_{A,C,B}+\ax_{A,D,B}}
\\
A(B(C+ D))          \ar[u]^{1_A\x\delta_{B,C,D}}
&&(AC)B+(AD)B      \ar@{<-}[d]^{\gx_{B,AC}+\gx_{B,AD}}
\\
A((C+ D)B)          \ar@{<-}[u]^{1_A\x\gx_{B,C+ D}}
                   \ar@{-->}[ruu]_{1_A\x\delta^\#_{C,D,B}}
&&B(AC)+ B(AD)      \ar@{<-}[d]^{\delta_{B,AC,AD}}    
\\
(A(C+ D))B         \ar[u]^{\ax_{A,C+ D,B}} 
&((AC)+(AD))B     \ar@{<-}@/^2pc/[l]^{\delta_{A,C,D}\x1_B}
                  \ar@{-->}[ruu]^{\delta^\#_{AC,AD,B}}
&B((AC)+(AD))     \ar@/^2pc/[l]^{\gx_{A,(AC)+(AD)}}
}\]
\caption{Laplaza Cond.~VIII}
\end{figure}
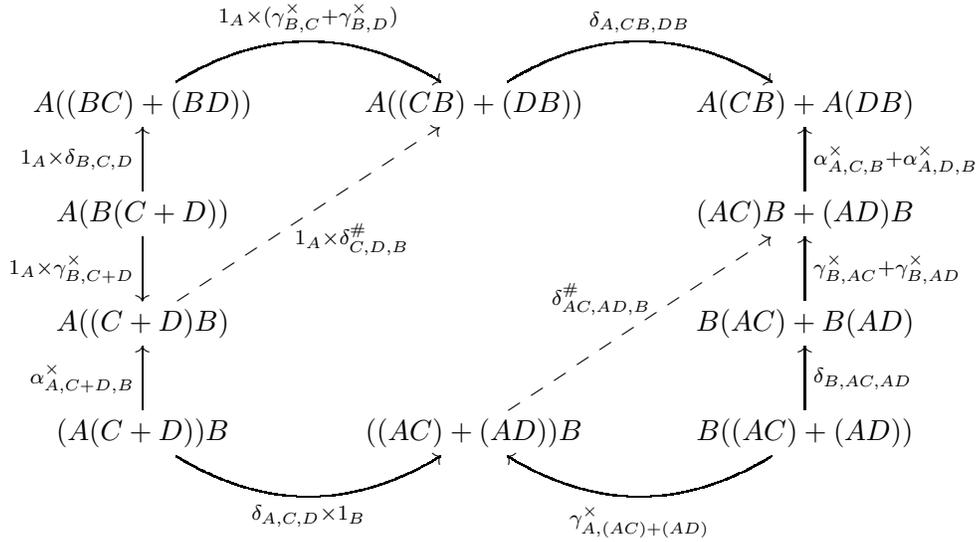
\noindent(As indicated in the captions, these results are two of
Laplaza's coherence conditions in \cite{laplaza}.)
As before, the dashed arrows represent right distributivity witnesses
$\delta^\#$.  The inner parts of these two figures, between the dashed
lines, give the desired simplifications of $\alpha^\times$ whenever
one of the subscripts is a sum.  So we need only define
$\alpha^\times$ when its subscripts are among the $\h x_i$'s. 

We can also eliminate the case where one of the subscripts is $\h
x_0=1$.  Figure~8 handles the case where the second subscript is 1:
\[
\alpha^\times_{A,1,B}=(\rho^\times_A\times1_B)*
(1_A\times\lambda^\times_B)^{-1}.
\]
The cases where the first or third subscript of $\alpha^\times$ is 1
are handled by Proposition~1.1 of \cite{js2} (a consequence of the
coherence conditions for braided monoidal structure, Figures~5--8)): 
\begin{align*}
\alpha^\times_{1,A,B}&=(\lambda^\times_A\times1_B)*
(\lambda^\times_{A\times B})^{-1}   \\
\alpha^\times_{A,B,1}&=(\rho^\times_{A\times B})*
(1_A\times\rho^\times_B)^{-1}.
\end{align*}
So $\alpha^\times$ needs to be defined only when the subscripts are
among the $\h x_i$'s for $1\leq i\leq q$.  

\subsection{Summary}
We have explcitly defined most of the structure of a unitary fusion
rig (based on given fusion rules). Specifically, we have defined the
raw elements, the witnesses, the operations on witnesses that provide
a witness frame, the binary operations $+$ and $\times$ and nullary
operations $0$ and $1$ on raw elements and on witnesses, and all of
the rig witnesses except $\alpha^\times$ and $\gamma^\times$. Even
these two exceptional cases have been reduced to the cases where the
subscripts are among the generators other than 1, by using the
formulas above for the cases where a subscript is a sum or is 1. (The
degenerate sum 0 of no summands is also covered, because any
$\alpha^\times$ or $\gamma^\times$ with a 0 subscript is a witness for
$0=0$ and is therefore uniquely determined.)  Furthermore, these
definitions and reductions ensure that all of the coherence conditions
except possibly Figures~5, 6, and 7 are satisfied.

This completes our description of unitary fusion rigs in
general. Specific unitary fusion rigs, for specific fusion rules, are
determined by witnesses $\alpha^\times_{\h x_i,\h x_j,\h x_k}$ and
$\gamma^\times_{\h x_i,\h x_j}$ with $i,j,k$ ranging from 1 to
$q$. Since we are dealing with finite tuples of finite-dimensional
Hilbert spaces, the data needed to specify a unitary fusion rig are
just finitely many matrices of complex numbers. (The matrices for
$\alpha^\times$ and $\gamma^\times$ are often called $F$-matrices and
braiding matrices.)  Unitary fusion rigs correspond to such systems of
matrices, subject to the requirements that the pentagon condition
(Figure~5) and the two hexagon conditions (Figures~6 and 7) must be
satisfied. 

For a given set of fusion rules, these coherence conditions may have
several solutions, or just one, or none. Accordingly, there may be
several unitary fusion rigs, or just one, or none for those fusion
rules.
 
In the next section, we review the calculation, done in detail in
\cite{G225}, for one particular fusion rule. The results for several
other fusion rules are given in Bonderson's thesis \cite{Bonderson},
along with indications of how to do some such computations more
efficiently on a computer.

\section{Fibonacci  Anyons}             \label{sec:fibonacci}

In this section, we show how a particular anyon model, Fibonacci
anyons, looks in our witness algebra framework.  In particular, we show
how the new framework accommodates the computations in \cite{G225}
of the associativity and braiding operators of this model.

The Fibonacci anyon model, already discussed in Section~\ref{fib1}, has
only two types, i.e., two generators of the true algebra, the
multiplicative identity $1$ (corresponding physically to the vacuum)
and the anyon type $\tau$.  The fusion rules are
$\tau\times \tau=\tau+1$ and the rules saying that 1 is a two-sided
identity element.

Applying the framework of unitary fusion rigs from the preceding
section, we have a raw algebra consisting of pairs of Hilbert spaces,
equipped with specified bases.  We prefer to write these pairs as
$(H_1,H_\tau)$ rather than $(H_0,H_1)$ because the types, rather than
their indices, are more informative subscripts.  Witnesses are pairs
of unitary transformations.  Addition is componentwise direct sum, and
multiplication is given by the sum-of-tensor-products formula as
above.

We intend to set up the computation of associativity and commutativity
witnesses in the unitary fusion semiring for Fibonacci anyons.  As
noted earlier, these witnesses in any fusion semiring are completely
determined by a rather small number of them, namely those with types
other than 1 as subscripts.  In the case of Fibonacci anyons, this
means that the fusion semiring structure will be completely determined
if we can compute two witnesses, $\alpha^\times_{\tau,\tau,\tau}$ and
$\gamma^\times_{\tau,\tau}$.  These witnesses are constrained by the
pentagon and hexagon conditions from the definition of braid
semirings, Figures~5--7.

Before proceeding to set up these computations, we simplify the
notation for the relevant basis elements, not only for the sake of
simplicity but also to match the notation that we used in
\cite{G225}.  

The raw element 1 (also known as $\h x_0$) is the pair of Hilbert
spaces $(\bbb C,0)$, and the specified basis of the first component
consists of the complex number 1.  We do not attempt to simplify this.

The raw element $\tau$ (also known as $\h x_1$) is $(0,\bbb C)$, and the specified basis
of the second component again consists of the complex number 1.  It
will be convenient to give this basis element a different name, so
that the name tells us where the basis vector came from.  We give it
the name $\tau$.  Thus, $\tau$ denotes the unique basis vector in the
raw element $\tau$, just as, in the preceding paragraph, 1 was the
unique basis vector in the raw element 1.

We shall rename some more complicated basis elements, in iterated
products of $\tau$'s, but we shall do so in a way that always allows
us to read off, from a basis element, the raw element that it
is in, and indeed the specific component of that raw element that it is in.

In fact, a suitable system of simplified names for basis elements in
iterated products of $\tau$, can be produced by a remarkably simple
scheme.  Begin with the basis vectors for the trivial products 1 and
$\tau$ as described above.  For non-trivial products $x\times y$, we
have, from the preceding section, the official notation
$(a,b,i,j,1)_k$, where $k$ tells us which component of $x\times y$
this basis vector is in, where $i$ and $j$ tell us which components of
the factors $x$ and $y$ the basis vectors $a$ and $b$ are taken from,
and where the final 1 tells us nothing because Fibonacci anyons have
all $N^k_{i,j}\leq 1$.  Instead of this official notation, we shall
use
\[
(a\underset{k}{\cdot}b),
\]
and we shall usually write $1$ or $\tau$ under the dot, instead of 0
or 1, respectively.  This shorter notation works because the
additional information $i,j$ in the official notation and the sources
$x,y$ can all be read off from $a$ and $b$.

\begin{ex}
  The simplest non-trivial product, $\tau\times\tau$, is a pair of
  1-dimensional Hilbert spaces.  The specified basis vector in the
  1-component (the first Hilbert space in the pair) is officially
  $(\tau,\tau,1,1,1)_0$, and our simplified notation for it is
  $\one\tau\tau$.

  Similarly, the specified basis vector in the $\tau$-component (the
  second Hilbert space in the pair) is officially
  $(\tau,\tau,1,1,1)_1$, and our simplified notation for it is
  $\two\tau\tau$.

In $(\tau\times\tau)\times\tau$, the first component is a
one-dimensional Hilbert space with specified basis vector
$\one{\two\tau\tau}\tau$.  The second component is two-dimensional
with specified basis vectors $\two{\one\tau\tau}\tau$ and
$\two{\two\tau\tau}\tau$.

Recalling that multiplication in the raw algebra is not generally
asociative, we also consider $\tau\times(\tau\times\tau)$.  Its
components have the same dimensions as in the preceding paragraph, but
the specified bases are potentially different.  The first component
has the basis vector $\one\tau{\two\tau\tau}$.  The second component
has the basis vectors $\two\tau{\one\tau\tau}$ and
$\two\tau{\two\tau\tau}$. 

Note that the parentheses in the simplified notation are in the same
places as the parentheses in the raw element
$(\tau\times\tau)\times\tau$ or $\tau\times(\tau\times\tau)$. \textqed 
\end{ex}

Continuing as in the example, we have simplified notations for all the
basis vectors in iterated products of $\tau$'s.

We are now in a position to discuss the associativity witness
$\alpha^\times_{\tau,\tau,\tau}$.  It consists of two unitary
transformations, one between the one-dimensional first components of
$(\tau\times\tau)\times\tau$ and $\tau\times(\tau\times\tau)$, and one
between their two-dimensional second components.  

For purposes of calculation, we want to exhibit these unitary
transformations as unitary matrices.  Exhibiting these matrices on the
page requires choosing an ordering of the vectors within each of the
specified bases for the two-dimensional spaces.  We arbitrarily order
them in the order in which they were mentioned above.  Now the two
components of $\alpha^\times_{\tau,\tau,\tau}$ are a unitary
$1\times1$ matrix (i.e., a complex number of absolute value 1) and a
$2\times 2$ unitary matrix, say
\[
(p) \qquad\text{and}\qquad
\begin{pmatrix}
  q&r\\s&t
\end{pmatrix}.
\]  
In detail, this means that 
\begin{align*}
\alpha^\times_{\tau,\tau,\tau}\one{\two\tau\tau}\tau &=p\one\tau{\two\tau\tau}  \\
\alpha^\times_{\tau,\tau,\tau}\two{\one\tau\tau}\tau &=
q\two\tau{\one\tau\tau}  +r\two\tau{\two\tau\tau}   \\
\alpha^\times_{\tau,\tau,\tau}\two{\two\tau\tau}\tau &=  
s\two\tau{\one\tau\tau}  + t\two\tau{\two\tau\tau}.
\end{align*}

These equations match those in \cite[Section~5.4]{G225} except that
we have here written the associativity witness
$\alpha^\times_{\tau,\tau,\tau}$ explicitly whereas in \cite{G225} the
corresponding isomorphism was used to identify the corresponding
components of  $(\tau\times\tau)\times\tau$ and
$\tau\times(\tau\times\tau)$ (see footnote~8 of \cite{G225}).  

The commutativity witnesses are similar but simpler.  Since both
components of $\tau\times\tau$ are one-dimensional,
$\gamma^\times_{\tau,\tau}$ is just a pair of $1\times1$ unitary matrices,
i.e., a pair of complex numbers $a,b$ of absolute value 1.  In detail,
\begin{align*}
  \gamma^\times_{\tau,\tau}\one\tau\tau   &=a \one\tau\tau   \\
\gamma^\times_{\tau,\tau} \two\tau\tau  &= b\two\tau\tau   .
\end{align*}
Again, these equations match those in \cite[Section~5.5]{G225}.

With this notation in place, and remembering that symmetry and
transitivity operations on witnesses are given by inversion and
composition of unitary transformations, it is easy to write out the
pentagon and hexagon conditions as equations for the matrix entries
$p,q,r,s,t,a,b$.  Since the computations are done in detail in
\cite{G225}, we do not repeat them here but merely record that the
pentagon condition reads
\[
  \begin{pmatrix}
    rs&q&rt\\
q&0&r\\st&s&t^2
  \end{pmatrix}=
  \begin{pmatrix}
    p^2q&prs&prt\\
prs&q^2+rst&qr+rt^2\\
pst&qs+st^2&rs+t^3
  \end{pmatrix}
\]
for the $\tau$ component and 
\[
  \begin{pmatrix}
    1&0\\
0&p^2
  \end{pmatrix}=
  \begin{pmatrix}
    q^2+prs&qr+ptr\\
qs+pts&rs+pt^2
  \end{pmatrix}
\]
for the 1 component.

It follows from these equations (in fact, from just the $\tau$
component --- the 1 component is redundant) and the unitarity of the
witness matrices that
\[
p=1,\qquad q=-t=\frac{-1+\sqrt5}2, \qquad r=\sqrt qe^{i\theta}, \qquad
s=\sqrt qe^{-i\theta}
\]
for some real $\theta$.  If we modify the standard basis vectors by
suitable phase factors, we can simplify the result by getting rid of
$\theta$, so $r=s=\sqrt q$.

The hexagon condition reads, if we take into account that $p=1$,
\[
  \begin{pmatrix}
    q^2+brs&(q+bt)r\\
(q+bt)s&rs+bt^2
  \end{pmatrix}=
  \begin{pmatrix}
    b^4q&b^3r\\
b^3s&b^2t
  \end{pmatrix}
\]
for the $\tau$ component and 
\[
a=b^2
\]
for the 1 component.

These equations and unitarity yield that, except for possible complex
conjugation of both $a$ and $b$,
\[
b=\frac{-1+\sqrt{q^2-4}}2=e^{3\pi i/5},\qquad
a=b^2=e^{6\pi i/5}.
\]

\end{document}